\documentclass[11pt, oneside,reqno]{amsart}   	
\usepackage[margin=1.5cm]{geometry}                		
\usepackage{amsaddr} \pdfoutput=1
 \usepackage{hyperref}
\usepackage{graphicx}			
\usepackage{amsmath}
\usepackage{amssymb}
\usepackage{xfrac}
\usepackage{color}
\usepackage{mathdots}
\usepackage{float}
\usepackage{cite}
\usepackage{mathtools}
\usepackage{tikz}
\usetikzlibrary{arrows}
 \newcommand{\rmi}{\mathrm{i}}
 \newcommand{\rme}{\mathrm{e}}
 \newcommand{\tr}{\mathrm{tr}}
 \newcommand{\Id}{\mathbb{I}}
 \hypersetup{
    colorlinks=true,
    linkcolor=black,
    citecolor=black,
    filecolor=black,
    urlcolor=blue,
}
\begin{document}

\title[]{Integrable spin chains and the Clifford group}
\author{Nick G. Jones}
\address{Mathematical Institute, University of Oxford, Oxford, OX2 6GG, UK\\
and the Heilbronn Institute for Mathematical Research, Bristol, UK}
\email{nick.jones@maths.ox.ac.uk}
\author{N. Linden}
\address{School of Mathematics, University of Bristol, Bristol BS8 1UG, UK}
\email{n.linden@bristol.ac.uk}

\begin{abstract} 
We construct new families of spin chain Hamiltonians that are local, integrable and translationally invariant. To do so, we make use of the Clifford group that arises in quantum information theory. We consider translation invariant Clifford group transformations that can be described by matrix product operators (MPOs). We classify the translation invariant Clifford group transformations that consist of a shift operator and an MPO of bond dimension two---this includes transformations that preserve locality of all Hamiltonians; as well as those that lead to non-local images of particular operators but nevertheless preserve locality of certain Hamiltonians. We characterise the translation invariant Clifford group transformations that take single-site Pauli operators to local operators on at most five sites---examples of Quantum Cellular Automata---leading to a discrete family of Hamiltonians that are equivalent to the canonical XXZ model under such transformations. For spin chains solvable by algebraic Bethe Ansatz, we explain how conjugating by a matrix product operator affects the underlying integrable structure. This allows us to relate our results to the usual classifications of integrable Hamiltonians. We also treat the case of spin chains solvable by free fermions.
\end{abstract}
\maketitle
\section{Introduction}
Integrable models are fundamental tools in our understanding of quantum many body systems and statistical mechanics \cite{Franchini,Korepin1997,Baxter2016,Guan13}. While a general definition of quantum integrability is open to debate \cite{Caux}, there are many examples of well-established integrable models. Two families of integrable models that are particularly well studied, and that are of interest to us, are Bethe Ansatz solvable models \cite{Korepin1997}, including the Heisenberg and XXZ spin chains \cite{Franchini,Korepin1997,Murg,Baxter2016,Sachdev01} and free-fermion models, including the XY and Ising spin chains \cite{Lieb61}. An important topic is the classification of Bethe Ansatz solvable models \cite{Kulish1982,Izergin1984,deLeeuw2019,deLeeuw2020,Gombor21,Ryan22}, and there are conjectured conditions implying integrability of a Hamiltonian \cite{Grabowski,Gombor21}. There has moreover been significant recent interest in finding new models and deriving conditions where a spin chain can be solved by free-fermion (and free-parafermion) methods \cite{Minami16,Minami17,Fendley2019,Chapman2020,Ogura2020,Yanagihara20,Alcaraz20,Alcaraz20b,Elman21}.

Given known integrable Hamiltonians, we can consider transformations applied to them and between them. This can lead to finding new models that can be solved in the same way as known models. Transformations between exactly solvable models are moreover important in the field of symmetry-protected topological (SPT) phases; then it is useful to identify transformations that take you from a simple, exactly solvable, model in the trivial phase to an analogous exactly solvable model with non-trivial symmetry properties \cite{Chen14,Verresen17,Scaffidi17}. In this paper we will construct new families of spin chain Hamiltonians using \emph{Clifford group transformations}, a family of transformations arising in quantum information theory that we describe below.

To set the scene, consider the following Hamiltonian for a spin-\sfrac{1}{2} chain:
\small 
\begin{align}
H_0 = -\sum_{n \in \textrm{sites}} \Big( Z_{n-3} Y_{n-2}X_{n-1}Y_{n}Y_{n+1}X_{n+2}Y_{n+3}Z_{n+4}+Z_{n-3} Y_{n-2}Y_{n-1}Y_{n+2}Y_{n+3}Z_{n+4}+  \Delta Z_{n-1}Y_{n}Y_{n+1} Z_{n+2}\Big),\label{eq:H_0}
\end{align}\normalsize
with $-1<\Delta<1$; and where $X$, $Y$ and $Z$ are the standard Pauli operators. Perhaps surprisingly, this Hamiltonian is integrable. In particular it can be analysed using Bethe Ansatz methods, and, from that, one can derive highly non-trivial results. For example, letting angle brackets denote the ground state expectation value, one can rigorously prove the following large $M$ asymptotics:
\begin{align}
\langle Z_1X_2Z_3 &~ Z_{M+1}X_{M+2}Z_{M+3} \rangle \\&= \begin{cases}-
\frac{1}{8\eta^2 M^2}(1+ o(1)) \qquad &0<\eta<\pi/4 \\
(-1)^M\frac{b}{M^{\frac{\pi}{2\eta}}}(1+ o(1))  \qquad &\pi/4<\eta<\pi/2,
\end{cases}\nonumber
\end{align} 
where $2\eta = \arccos(\Delta)$ and $b$ is independent of $M$. As we will explain, this integrability follows by giving a unitary transformation from $H_0$ to the canonical XXZ spin chain
\begin{align}
H_\textrm{XXZ} = -\sum_{n \in \textrm{sites}}\Big(  X_{n}X_{n+1}+Y_{n}Y_{n+1} +  \Delta Z_{n}Z_{n+1}\Big); \label{eq:H_XXZ}
\end{align}
 we can then use established results for the $\langle Z_1Z_{M+1}\rangle$ correlation function in the XXZ chain \cite{Boguliubov1986}. Moreover, using conformal field theory (CFT) methods ($H_0$ has low energy behaviour described by a compactified free boson CFT for $-1<\Delta<1$, with SU(2)-point at $\Delta=-1$), one can derive further ground state correlation functions in this model \cite{Boguliubov1986,Korepin1997,diFrancesco1999}. We will show that, unlike $H_{\mathrm{XXZ}}$, $H_0$ is at a transition between SPT phases \cite{Pollmann10,Chen11,Schuch11,Verresen17} (indeed, unitary transformations preserve the structure of the phase diagram, but not necessarily any symmetry properties). 
 
Now, the fact that, say, the eigenstates of the Hamiltonians $H$ and $H' = UHU^\dagger$, for some unitary $U$, are simply related is elementary. However, a typical unitary will not preserve the locality of the Hamiltonian; by which we mean that the Hamiltonian is a sum of terms, each of which acts non-trivially on a number of sites that is independent of the length of the chain. Moreover, a given local Hamiltonian is typically not integrable. The purpose of this paper is to construct new translationally invariant and local Hamiltonians that are integrable by virtue of being unitarily related to canonical integrable models, such as $H_\textrm{XXZ}$. We do this through making connections to techniques from quantum information theory. In particular, we study interesting examples from the \emph{Clifford group} of unitary transformations. 

The Clifford group is a discrete subgroup of the full unitary group and consists of those transformations that map products (or strings) of Pauli operators to other products of Pauli operators \emph{under conjugation}; for example:
\begin{align} X_1\rightarrow X_1 \qquad &X_{2}\rightarrow Z_{2}\nonumber\\
Z_1\rightarrow Z_1Z_{2} \qquad &Z_{2}\rightarrow X_1X_{2} \label{eq:exampletransformation}
\end{align} 
is a two-site Clifford group transformation. Since these transformations are unitary, they are constrained by the necessity of preserving the commutation relations between Pauli operators. One reason for considering the Clifford group is that many integrable spin-$\sfrac{1}{2}$ chains have a simple form written as a sum of Pauli strings and these transformations preserve that---for example, $H_\mathrm{XXZ}$ has three types of interaction term, as does $H_0$. Symmetries that are Pauli strings, such as the spin-flip $\prod_n Z_n$, remain of this form after the transformation. This can help in the analysis of symmetries in the context of SPT phases \cite{Pollmann2017,Verresen21}; and so can be used to find examples of integrable models (both gapped and gapless) with non-trivial SPT physics. For the case of $H_0$, this analysis is given in Appendix \ref{app:H0}.

Consider then a spin-$\sfrac{1}{2}$ chain with $L$ sites. One approach to finding Clifford group transformations on such a chain is to start with a simple transformation for each site, such as the following that appears frequently in the literature on SPT phases \cite{Chen14,Verresen17,Scaffidi17}:
\begin{align}
X_n &\rightarrow  Z_{n-1}X_nZ_{n+1}\nonumber\\ 
Z_n &\rightarrow Z_n. \label{eq:clustermapping}
\end{align} 
This transformation maps the trivial paramagnet with Hamiltonian $-\sum_n X_n$ to the SPT cluster model with Hamiltonian $-\sum_n Z_{n-1}X_nZ_{n+1}$ (for this reason it is sometimes referred to as an \emph{SPT entangler}). One can check that \eqref{eq:clustermapping} preserves the commutation relations, giving a Clifford group transformation\footnote{We ignore any boundary conditions here. In fact, accounting for the boundary in this transformation is one way to understand the edge modes characteristic of SPT phases in the cluster model \cite{Verresen17}.}. One nice feature of this transformation is that it preserves translation invariance---this is a physically interesting case and is a common, though inessential \cite{Boguliubov1986,Baxter2016}, feature of many integrable models. A key concern of this paper is to identify translation invariant transformations that in turn lead to translation invariant Hamiltonians. 

Note that the transformation \eqref{eq:clustermapping}, taking single-site operators to (at most) three-site operators for each site $n$, defines a Clifford group transformation on the whole chain. Let us consider instead the following partial definition of a transformation taking single-site operators to three-site operators:
\begin{align}
X_1 &\rightarrow  X_{0}X_1Z_{2}\nonumber\\ 
Z_1 &\rightarrow Z_1. \label{eq:notTIexample}
\end{align} 
There are many ways of extending this to a valid three-site Clifford transformation, and moreover to a Clifford transformation on the whole chain. However, any way we do so will break translation invariance. In particular, there cannot be a valid Clifford transformation on the whole chain that acts as \eqref{eq:notTIexample} on each site:
\begin{align}
X_n &\rightarrow_{?}  X_{n-1}X_nZ_{n+1}\nonumber\\ 
Z_n &\rightarrow_{?} Z_n. \label{eq:inconsistentexample}
\end{align} 
Indeed such a transformation would not be unitary since, say, mapping $X_1 \rightarrow  X_{0}X_1Z_{2}$ and $X_3  \rightarrow  X_{2}X_3Z_{4}$ is inconsistent with the commutation relation $[X_1,X_3]=0$. We thus see that it is a non-trivial problem to identify those local Clifford transformations that can apply in a translation invariant fashion on the whole chain. We will present families of such transformations below. 
\begin{figure}[ht]\scalebox{0.7}{
\includegraphics{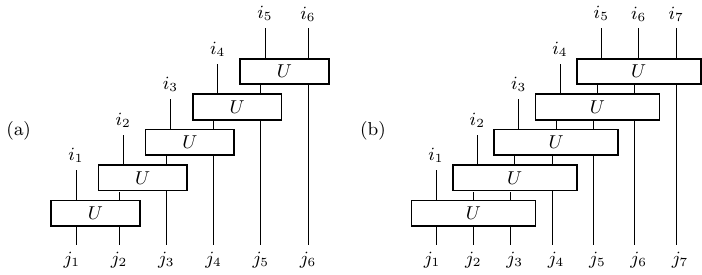}}
\caption{Translation invariant Clifford group transformations (a) $U_{\mathrm{chain}}  = \dots U_{3,4}U_{2,3}U_{1,2}$ and (b) $U_{\mathrm{chain}} = \dots U_{3,4,5}U_{2,3,4}U_{1,2,3}$.}
\label{fig:uchain}
\end{figure}

General Clifford transformations can be built from a discrete set of one and two-site operations: CNOT, Hadamard and Phase \cite{Ozols,Gottesman} with matrix forms
\begin{align}
\left(\begin{array}{cccc}1 & 0 & 0 & 0 \\0 & 1 & 0 & 0 \\0 & 0 & 0 & 1 \\0 & 0 & 1 & 0\end{array}\right),\qquad\frac{1}{\sqrt{2}}\left(\begin{array}{cc}1 & 1 \\1 & -1\end{array}\right) \quad \mathrm{and} \quad \left(\begin{array}{cc}1 & 0 \\0 & \rmi\end{array}\right)
\end{align}
respectively. It is natural then to consider how to construct a translation invariant Clifford group transformation out of basic local Clifford transformations. One way to extend a \emph{single} basic Clifford transformation $U$ to a Clifford transformation $U_\textrm{chain}$ acting on the whole chain is given in Figure \ref{fig:uchain} for two- and three-site $U$ respectively. In fact, the transformation \eqref{eq:clustermapping} takes this form with a basic Clifford transformation that acts by conjugation as: 
\begin{align}
 X_{n}\rightarrow X_{n}Z_{n+1} \qquad &  Z_{n}\rightarrow Z_{n} \nonumber\\
 X_{n+1}\rightarrow Z_{n}X_{n+1} \qquad &  Z_{n+1}\rightarrow Z_{n+1}\label{eq:cluster2site}.
\end{align}

While this construction allows us to build translation invariant Clifford transformations from simple basic Clifford transformations $U$, the transformations $U_\textrm{chain}$ of Figure \ref{fig:uchain} have a non-local causal structure, and so, for a general choice of $U$, will not necessarily preserve the locality of the Hamiltonian that we transform. Indeed, consider applying \eqref{eq:exampletransformation} sequentially to pairs of sites, as in Figure \ref{fig:uchain}(a): this means that $U_\textrm{chain}$ is a product of basic transformations that act by conjugation as: 
 \begin{align} X_n\rightarrow X_n \qquad &X_{n+1}\rightarrow Z_{n+1}\nonumber\\
Z_n\rightarrow Z_nZ_{n+1} \qquad &Z_{n+1}\rightarrow X_nX_{n+1}.\label{eq:KWtransform}
\end{align} 
We can see that on applying $U_\textrm{chain}$, the operator $X_2$, for example, will transform into a non-local string: $X_2\rightarrow Z_2\rightarrow Z_2Z_3\rightarrow Z_2Z_3Z_4 \rightarrow \dots$. One part of our analysis is identifying which cases maintain locality for all choices of Hamiltonian. We will call these transformations, that take all local operators to other local operators, \emph{locality-preserving} transformations (where a local operator acts non-trivially on a number of sites independent of the length of the chain). The other Clifford transformations, that we will call \emph{locality-non-preserving}, are, however, of real physical interest. In particular, while they take some local operators to infinite strings, they can transform certain local Hamiltonians to other local Hamiltonians (in the bulk). For example, the operator $U_\textrm{chain}$ derived from \eqref{eq:KWtransform} can be recognised as the Kramers-Wannier duality \cite{Kramers41} of the Ising model (again, for operators in the bulk; we will discuss this further below, see also \cite{Aasen16}).
 
On a spin-\sfrac{1}{2} chain, transformations that map local strings of Pauli operators to local strings of Pauli operators are locality-preserving. In fact, the purely mathematical question that we arrive at by asking which translation invariant Clifford transformations maintain locality for all Pauli operators arises in the field of Clifford \emph{Quantum Cellular Automata} (QCAs). To be precise, a Clifford QCA is a translation invariant, reversible and local transformation that takes Pauli operators to Pauli operators \cite{Schumacher2004,Schlingemann2008}. The motivation in that setting is rather different: a QCA is a discrete time dynamical system, and the purpose of the locality condition is to ensure that the system has a finite propagation speed. There is an established theory of Clifford QCAs \cite{Schumacher2004,Schlingemann2008,Farrelly2020} and certain claims made below are consequences of these more general results---we will point these out as appropriate, but aim to give a self-contained presentation and so provide elementary proofs as needed. Moreover, the condition that locality is preserved for every Hamiltonian is stronger than needed for our purposes. Allowing for locality-non-preserving transformations, that nevertheless preserve the locality of certain Hamiltonians, leads to a broader class of transformations than those that define a QCA.

Another important connection to make is to the theory of matrix product operators (MPOs) \cite{Verstraete2004,Zeng2019}. We explain in Section \ref{sec:TIClifford} that the transformation $U_\textrm{chain}$ in Figure \ref{fig:uchain} can be understood as an MPO. This is particularly useful because both MPOs and the Algebraic Bethe Ansatz \cite{Korepin1997,Franchini} can be formulated in the intuitive graphical notation used in the theory of tensor networks \cite{Murg}. This allows us to begin to analyse, for models solved by Algebraic Bethe Ansatz, how the integrable structure changes when we conjugate the Hamiltonian by the transformation $U_\textrm{chain}$.

The paper is organised as follows. In Section \ref{sec:Clifford} we motivate and analyse certain families of Clifford group transformations. In Sections \ref{sec:Cliffordgroup} and \ref{sec:TIClifford} we define Clifford group transformations in general, and then the translation invariant Clifford group transformations of interest. We also explain the connection to MPOs. In Sections \ref{sec:XXZfamilies} and \ref{sec:generalisations} we then demonstrate how these results can be used to construct families of integrable models equivalent to $H_\textrm{XXZ}$, including $H_0$ given above. Models related by locality-preserving transformations built from a single two-site basic Clifford transformation are given in Section \ref{sec:localXXZ}; while those related by locality-non-preserving transformations are given in Section \ref{sec:nonlocal}. 
These are a consequence of a classification, given in Section \ref{sec:twositeClifford}, of all of the possible transformations that arise from transformations $U_\textrm{chain}$ constructed from a single two-site Clifford transformation. We see that, up to changes of on-site basis, only one new Hamiltonian appears in Section \ref{sec:localXXZ}. In Section \ref{sec:folded} we show that this Hamiltonian has some interesting connections to another integrable model, the \emph{folded XXZ chain} \cite{Yang20,Zadnik21,Zadnik21b,Pozsgay21,Gombor21}, and to the recently introduced concept of a pivot Hamiltonian \cite{Tantivasadakarn2021,Tantivasadakarn2021b}. In Section \ref{sec:generalisations} we discuss translation invariant Clifford transformations directly; rather than through the $U_\mathrm{chain}$ construction. We give, in Section \ref{sec:generalisationsresults}, all transformations that map single-site Pauli operators to local operators on up to five sites and show how these can be written using a $U_\textrm{chain}$ based on a single two- or three-site Clifford transformation. We give the corresponding families of Hamiltonians related to $H_{\textrm{XXZ}}$ in Section \ref{sec:summary}.
In Section \ref{sec:integrable}, we proceed to a more general discussion regarding families of integrable Hamiltonians. We focus our discussion on canonical models that can be solved by the quantum inverse scattering method, or algebraic Bethe Ansatz \cite{Korepin1997,Franchini,Murg}, and explain how transformations of MPO form can be applied to the integrable structures underlying these models. This includes, as a special case, the Clifford transformations that we have analysed and thus gives our results a context by relating them to classifications of integrable models. We find interesting modifications to the underlying integrable structures, but there remain questions to resolve here.
 We also treat the case of spin chains solvable by transforming to free fermions, making connections to recent papers on families of exactly solvable spin-$\sfrac{1}{2}$ chains \cite{Minami17,Chapman2020,Ogura2020,Elman21}. Finally, we discuss further avenues of research.  
\section{Clifford group transformations}\label{sec:Clifford}
 \subsection{The Clifford group}\label{sec:Cliffordgroup}
 We now introduce the Clifford group in the spirit of \cite{Ozols}. Let $\mathcal{P}=\{ \Id, X, Y,Z\}$ denote the set of single-site Pauli operators. Consider the set of Pauli strings on $N$ sites, this is given by:
 \begin{align}
 \mathcal{P}_N = \big\{ P_1 \otimes P_2 \dots \otimes P_N \big\vert P_j \in \mathcal{P} \big\}. 
 \end{align}
 As above, we will omit the tensor product symbol and write elements of $ \mathcal{P}_N$ as $\prod_{j=1}^N P_j$.  Furthermore, non-trivial Pauli strings on $N$ sites are elements of     \begin{align}
 \mathcal{P}^*_N  =\mathcal{P}_N\setminus \{\Id^{\otimes n} \} . \end{align} As mentioned, informally the Clifford group consists of transformations that take Pauli strings to Pauli strings under conjugation. More carefully, the Clifford group on $N$ sites is defined as the following subgroup of $U(2^N)$:
  \begin{align}
 \mathcal{C}_N  = \big\{ U \in U(2^N) \big\vert \forall P \in \pm\mathcal{P}^*_N, ~  UPU^\dagger \in \pm\mathcal{P}^*_N  \big\}/U(1),\end{align}
 where $\pm\mathcal{P}^*_N$ consists of $\pm P$ for $P \in \mathcal{P}^*_N$.

  One can specify\footnote{If we know how each of these operators transform, we can use $U \left(\prod_{j=1}^NP_j\right) U^\dagger = \prod_{j=1}^N (U P_jU^\dagger)$, linearity and that $Z_jX_j = \rmi Y_j$ to find the action on any element of $\pm\mathcal{P}^*_N$.} a Clifford group transformation by its action on each of $\{X_1,\dots,X_N, Z_1,\dots, Z_N\}$. These operators all mutually commute, apart from operators on the same site where we have the anticommutator $\{X_j,Z_j\}=0$. These commutation relations are preserved under conjugation by unitary transformations, and in particular by Clifford group transformations. Respecting these commutation relations, one can define a Clifford group transformation sequentially, first specifying, say, $X_1 \rightarrow P$, where $P \in  \pm\mathcal{P}^*_N$, then defining $Z_1 \rightarrow P' $, where $P' \in  \pm\mathcal{P}^*_N$ and $\{P,P'\}=0$, and so on.
 It can be shown that this procedure gives a unique element of the Clifford group \cite{Ozols} and so, in this way, one can derive the order of $\mathcal{C}_N$. For example, to define a transformation in $\mathcal{C}_1$, one can choose to take $X$ to any of the six elements of $\pm\mathcal{P}^*_1$, and $Z$ to any of the four elements of  $\pm\mathcal{P}^*_1$ that anticommute with that choice, giving $\lvert\mathcal{C}_1\rvert = 24$. The result $\lvert\mathcal{C}_N\rvert = 2^{N^2+2N}\prod_{j=1}^N(4^j-1)$ can be derived similarly, as shown in \cite{Ozols}. For our spin chain of length $L$ we will consider many copies of, say, $\mathcal{P}_2$. These will correspond to each pair (or, more generally in the case of $\mathcal{P}_{N>2}$, set) of neighbouring sites. For example, we can view both $X_1X_2$ and $X_2X_3$ as elements of $\mathcal{P}_2$, for different subsets of sites on the chain.

\begin{figure}[t]
\scalebox{0.75}{
\includegraphics{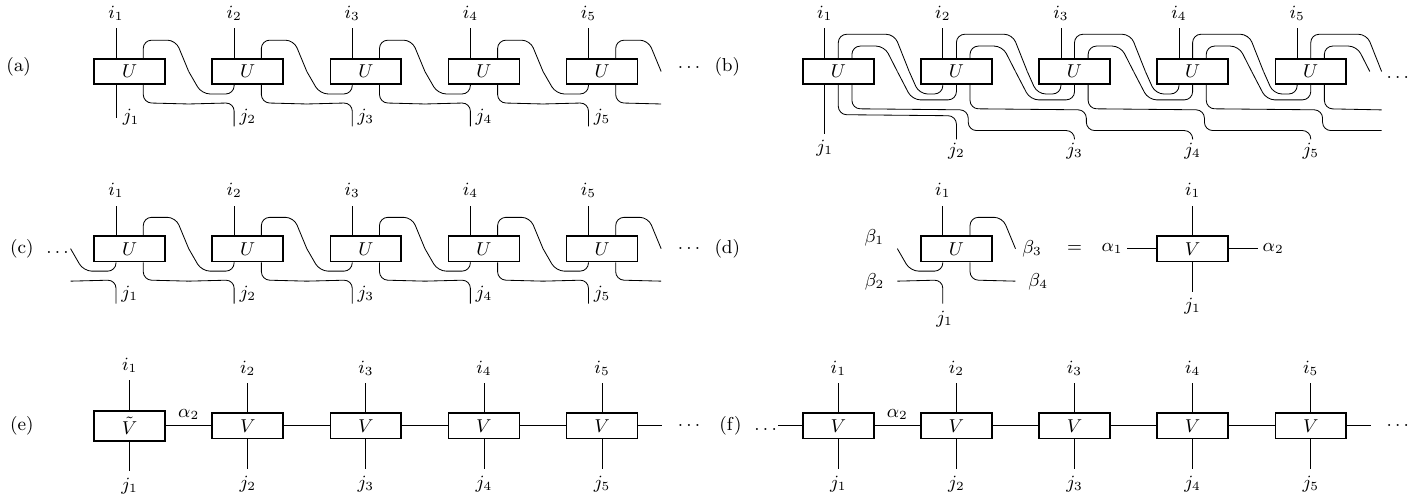}}
\caption{Relating the translation invariant Clifford group transformations of Figure \ref{fig:uchain} and MPOs. Figure (a) represents $U_{\mathrm{chain}}  = \dots U_{3,4}U_{2,3}U_{1,2}$ and Figure (b) represents  $U_{\mathrm{chain}}  =  \dots U_{3,4,5}U_{2,3,4}U_{1,2,3} $. Figure (c) is a  periodic analogue of Figure (a) given by $U_{\mathrm{chain}} =\dots U_{3,4}U_{2,3}U_{1,2}\dots$ this is interpreted in the main text. In Figures (a), (b), (c) all lines represent two-dimensional spaces. Figure (d) defines an MPO tensor $V$ in terms of the $U$ that appears in Figures (a) and (c). Figure (e) represents an MPO with open boundary, the virtual index $\alpha$ has dimension $\chi$. Figure (f) represents an MPO with periodic boundary, the virtual index $\alpha$ has dimension $\chi$. This shows that Figures (a) and (c) are of MPO form with $\chi =4$.} 
\label{fig:MPOuchain}
\end{figure}
 
 We define notation that corresponds to translating a Pauli operator by $j$ sites: if $ P = P^{(1)}_{1} P^{(2)}_{2} \dots  P^{(n)}_{n} \in\mathcal{P}_N $, then let $P(j) = P^{(1)}_{j+1} P^{(2)}_{j+2} \dots P^{(n)}_{j+N}$, with $P^{(k)}\in \mathcal{P}$ fixed.  Given a transformation in $\mathcal{C}_N$, we specify the sites it acts on and so we can then find the appropriate action on $P(j)$. To illustrate this, let $U\in \mathcal{C}_2$ be defined by:
$X_1 \rightarrow X_1 X_2$, $X_2 \rightarrow Z_1 Z_2$, $Z_1 \rightarrow Z_1$ and $Z_2 \rightarrow  X_2$. 
Then, for $Q = X_1 X_2$, we have: 
\begin{align} 
 U_{12}Q U^\dagger_{12}&= -Y_1Y_2\nonumber\\ 
U_{23}QU^\dagger_{23}& = X_1X_2X_3\nonumber \\
U_{j+1,j+2}Q(j)U^\dagger_{j+1,j+2}& = -Y_{j+1}Y_{j+2}.
\end{align}

Local Pauli operators are those that are elements of $\mathcal{P}_N$, where $N$ is independent of $L$, the length of the chain. The translation invariant spin chain Hamiltonians that we consider take the form:
\begin{align}\label{eq:generalhamiltonian}
H = \sum_{n} \sum_\alpha J_\alpha P^\alpha(n) \qquad J_\alpha \in \mathbb{R}, \qquad  P^\alpha \in \mathcal{P}_N,
\end{align}
where $\alpha$ labels the different terms that appear, and $n$ indexes the sites, where each site has a translated Pauli operator. The couplings $J_\alpha$ are all real since $H$ is Hermitian. For example, $H_0$ defined in \eqref{eq:H_0} would have $\alpha\in\{1,2,3\}$, $n \in \mathbb{Z}$ and $P^1 =Z_{1} Y_{2}X_{3}Y_{4}Y_{5}X_{6}Y_{7}Z_{8}$.

 \subsection{Translation invariant Clifford transformations and matrix product operators}\label{sec:TIClifford}
While we can look at transformations in $\mathcal{C}_L$ directly, we focus on those Clifford transformations that are of the form $U_\textrm{chain}$, illustrated in Figure \ref{fig:uchain}. This is given by:
\begin{align}
 U_{\mathrm{chain}} &= \prod_{n \in \textrm{sites}} U_{n,n+1,\dots,n+k-1} \qquad U \in \mathcal{C}_k ,\label{eq:uchainopen}\end{align}
 where the product is ordered so that we act first with the operator acting on the left-most sites of the chain. Such transformations are illustrated for $k=2$ and $k=3$ in Figure \ref{fig:uchain}. For $k=1$ this is an on-site change of basis. Note that throughout we are thinking of a large system size $L$, but will not concern ourselves with any subtleties arising in the thermodynamic limit $L\rightarrow \infty$ (see, for example, the rigorous approach in \cite{Schumacher2004}). 
Now, as mentioned in the introduction, this transformation has the form of an MPO, and this interpretation will be particularly helpful in Section \ref{sec:integrable} where we discuss connections to Bethe Ansatz methods. We use the standard tensor network graphical notation below \cite{Cirac21}, although we will also write out some formulae explicitly. In Figure \ref{fig:MPOuchain} we graphically relate the translation invariant Clifford group transformations of Figure \ref{fig:uchain} to MPOs.
This particular argument appears, for example, in the theory of matrix product states (MPS), showing equivalence between MPS and circuits of the form $U_\mathrm{chain}$ (not necessarily constructed from basic Clifford transformations) \cite{Schoen05,Smith20,Cirac21,Barratt21}. A translation invariant MPO on a chain $\bigotimes_{n=1}^L\mathcal{H}_n$ is an operator $M$ on the chain of the form \begin{align}
M_{j_1,\dots ,j_L}^{i_1,\dots,i_L}&=     \tr(V_{j_1}^{i_1}V_{j_2}^{i_2}\cdots V_{j_L}^{i_L})=\sum_{\alpha_1,\dots,\alpha_L}\left(V_{j_1}^{i_1}\right)^{\alpha_1}_{\alpha_2}\left(V_{j_2}^{i_2}\right)^{\alpha_2}_{\alpha_3}\cdots \left(V_{j_L}^{i_L}\right)^{\alpha_L}_{\alpha_1}.\label{eq:MPO}
 \end{align} For fixed $i_k,j_k$ (physical indices corresponding to $\mathcal{H}_k$), each of the $V_{j_k}^{i_k}$ are $\chi\times\chi$ matrices acting on a `virtual space' as $V_{j_k}^{i_k}: \mathcal{V}\rightarrow \mathcal{V}$, where virtual indices are then summed over (i.e., we matrix multiply in the virtual space) \cite{Verstraete2004,Zeng2019,Cirac21}. The virtual space $\mathcal{V}$ has dimension $\chi$, referred to as the bond dimension\footnote{Note that this need not coincide with the dimension of the physical sites.}. Alternatively, one can think of $V$ as a four index tensor, with two physical indices and two virtual indices. The trace over the virtual space, illustrated in Figure \ref{fig:MPOuchain}(f), reflects periodic boundary conditions (i.e., $\mathcal{H}_1$ and $\mathcal{H}_L$ are neighbouring sites). For open boundary conditions, we do not take the trace and can simply alter the tensors at the boundary (for fixed physical indices, the boundary tensors are vectors rather than matrices)---this is illustrated in Figure \ref{fig:MPOuchain}(e). To see that $U_\textrm{chain}$ defined in \eqref{eq:uchainopen} is an MPO, we show in Figure \ref{fig:MPOuchain}(d) how to find the MPO tensor $V$ from the basic Clifford transformation $U$. We see that for $k=2$ we have\footnote{In fact, $U_{\mathrm{chain}}$ with $k=2$ is the product of two $\chi=2$ MPOs. The left-moving shift operator, an MPO $M_{\textrm{shift}}$ with $\chi =2$ that takes site $n+1$ to site $n$ \cite{Cirac21}, and a $\chi =2$ MPO $M_U$ constructed from $U$ combine to give $U_\textrm{chain} = M_U M_{\textrm{shift}}$.} a $\chi=4$ MPO, and the generalisation for $k=3$ gives a $\chi=16$ MPO. 
 
 Now, if the operator $M_{j_1,\dots ,j_L}^{i_1,\dots,i_L}$ defined in \eqref{eq:MPO} is unitary, then it is called a matrix product unitary (MPU)---note that we take the trace and so this operator is of the periodic boundary condition form. It has been shown that MPUs are equivalent to QCAs \cite{Cirac2017,Csahinouglu2018,Piroli20}, and so will necessarily be locality-preserving. 
 
Let us now relate this equivalence to the MPOs considered in this work. Firstly, using the graphical argument of Figure \ref{fig:MPOuchain}, our definition of $U_\textrm{chain}$ in \eqref{eq:uchainopen} is an MPO with open boundary (there is no trace corresponding to summing over a virtual index between site $L$ and site $1$). Since the basic Clifford transformation $U$ is unitary, the resulting MPO is unitary and is moreover a Clifford transformation in $\mathcal{C}_L$. However, this does not define an MPU, since the definition of an MPU is for the periodic boundary condition form of the MPO. Hence, there is no equivalence between this construction\footnote{For a detailed discussion of the relationship between tensor networks and QCAs see \cite{Piroli20}.} and QCAs. Indeed, for $U_\mathrm{chain}$ derived from \eqref{eq:KWtransform}---the Kramers-Wannier duality---we saw that this family of MPOs is not necessarily locality-preserving. In fact, this duality is discussed in detail in \cite{Aasen16}. In that paper, a \emph{periodic} MPO form for the Kramers-Wannier duality is given. One can show that this MPO fits the construction we are considering, and is built out of a basic Clifford transformation\footnote{This basic Clifford transformation is the same as \eqref{eq:KWtransform} followed by an on-site change of basis on site $n$: $X_n\rightarrow Z_n$, $Z_n\rightarrow X_n$. In $U_\mathrm{chain}$ this means we do this basis change on every site apart from the right-most site.}:
\begin{align} X_n\rightarrow Z_n \qquad &X_{n+1}\rightarrow Z_nX_{n+1}\nonumber\\
Z_n\rightarrow X_nZ_{n+1} \qquad &Z_{n+1}\rightarrow Z_{n+1}.\label{eq:KWtransform2}
\end{align} 
Using this basic Clifford to construct $U_\mathrm{chain}$ leads to the (bulk) transformation $X_n \rightarrow Z_{n-1}Z_{n}$ and $Z_nZ_{n+1}\rightarrow X_n$. The corresponding periodic MPO (as in Figure \ref{fig:MPOuchain}(c)) is more subtle. The analysis in \cite{Aasen16} shows that it is not invertible (and hence not unitary) by analysing the fusion algebra with the spin-flip operator $\prod_n X_n$. More generally, if we take a basic Clifford transformation $U$ and use this to construct a periodic boundary MPO (given in Figure \ref{fig:MPOuchain}(c)), this is not necessarily unitary and so is not necessarily an MPU. However, if $U_\textrm{chain}$ is locality preserving, then conjugation by $U_{\mathrm{chain}}$ acts as a Clifford QCA. This means by the general arguments in \cite{Cirac2017,Csahinouglu2018} that there exists a periodic MPU form. Our analysis below allows us, using elementary calculations, to give such a periodic form directly for transformations of interest (see Section \ref{sec:commuting}). 

Our calculations focus on the open boundary case, which explicitly breaks translation invariance by fixing a left-most site. However, such transformations preserve bulk translation invariance in the following way. Let us index the sites here so that the first site has $n \simeq-L/2$ and the last site has $n \simeq L/2$. Then, given $P$, a local Pauli operator in the bulk, if $U_\textrm{chain}$ transforms $P\rightarrow Q$, where $Q$ is local, then we have that the translated operators transform as $P(j) \rightarrow Q(j)$ for $j$ independent of $L$. Note that the causal structure of $U_{\mathrm{chain}}$ means that we can have a locality-non-preserving transformation such that, on conjugating local operators, the non-local images extend to the right. Instead of conjugating by $U_\textrm{chain}$ we could conjugate by $U^\dagger_\textrm{chain}$, giving locality-non-preserving transformations where the non-local images extend to the left. We do not analyse this case, but analogous calculations would go through.

\section{Families of spin chains from the XXZ chain---\texorpdfstring{$U_{\mathrm{chain}}$ with $k=2$}{Uchain with k=2}}\label{sec:XXZfamilies}

In this section we characterise all local spin chains that are equivalent to $H_{\textrm{XXZ}}$ under conjugation by any $U_{\mathrm{chain}}$ that is built from a basic two-site Clifford transformation $U\in \mathcal{C}_2$. We first summarise the results for locality-preserving transformations in Section \ref{sec:localXXZ} and then summarise the results for locality-non-preserving transformations in Section \ref{sec:nonlocal}. We then prove our claims by classifying all $U_{\mathrm{chain}}$ with $k=2$ in Section \ref{sec:twositeClifford}. Of course, this classification does not depend on the Hamiltonian, so these results could be easily applied to other canonical models. For example, one could straightforwardly generalise the analysis to the integrable XYZ chain \cite{Baxter2016} simply by allowing independent couplings for each term in \eqref{eq:H_XXZ}. Note also that $H_{\textrm{XXZ}}$ is integrable for all $\Delta \in \mathbb{R}$. For $\Delta>1$ ($\Delta<-1$) the model is in a gapped (anti-)ferromagnetic phase, while for $\lvert\Delta\rvert<1$ we have a gapless paramagnetic phase \cite{Franchini}. If we include an external field term, $h\sum_n Z_n$, the model remains integrable. The phase diagram in the $\Delta-h$ plane has three regions corresponding to the three phases mentioned above for $h=0$; the value of $\Delta$ for the boundaries of these regions varies with $h$.

\subsection{Locality-preserving transformations with \texorpdfstring{$k=2$}{k = 2} and the XXZ chain}
\label{sec:localXXZ}
Consider all locality-preserving $U_\textrm{chain}$ built from a two-site Clifford transformation, as defined in \eqref{eq:uchainopen} for $k=2$ (see also Figure \ref{fig:uchain}(a)). In Section \ref{sec:twositeClifford} below, we will show that acting with any such locality-preserving transformation has the effect on $H_\textrm{XXZ}$ of either (a) performing an on-site change of basis, or (b) transforming
 $H_\textrm{XXZ}\rightarrow H'_{\textrm{XXZ}}$ where 
\begin{align}
{H}_{\textrm{XXZ}}' &= -\sum_{n\in\textrm{sites}} \Big(J_1 Z_{n-1}X_{n} X_{n+1}Z_{n+2} + J_2 Z_{n-1}Y_{n} Y_{n+1}Z_{n+2} + J_3 Z_nZ_{n+1}\Big),\label{eq:H'XXZ}
\end{align}
up to an on-site change of basis.
Here, two of the $J_i$ are equal to 1 and the third is equal to $\Delta$.  Note that the transformation acts as \eqref{eq:clustermapping} for this particular choice of on-site basis and for $J_3 = \Delta$.

A sum over four-spin interaction terms of the form $Z_{n-1}X_nX_{n+1}Z_{n+2}$ gives the Hamiltonian of a generalised cluster model \cite{Verresen17}. Such terms appear in models dual to free-fermion chains \cite{Suzuki71,Keating2004,Smacchia11,deGottardi13,Ohta16,Verresen2018}, through a standard Jordan-Wigner transformation (see the discussion in Section \ref{sec:free} below). However, applying this Jordan-Wigner transformation to ${H}_{\textrm{XXZ}}$ will lead to four-fermion interactions for $\Delta\neq0$. We also note that a model with these interaction terms appears in the literature as a `twisted' XXZ chain \cite{Scaffidi17,Parker,Duque21}. However, that model is not translation invariant and moreover contains additional interactions, meaning that it is not equivalent to ${H}_{\textrm{XXZ}}' $.

Note that $H_\textrm{XXZ}$ has a $U(1)$ symmetry corresponding to rotations about the $Z$-axis, and so this direction is distinguished. After the transformation, the distinguished direction corresponds to the term with $J_i = \Delta$. In all of the models above, we preserve integrability if we add an external field term of the form $\sum_n U^{\vphantom{\dagger}}_\mathrm{chain}Z_nU_\mathrm{chain}^\dagger$. For example:
\begin{align}
{\tilde H'_{\textrm{XXZ}}} &= -\sum_{n\in\textrm{sites}}\Big(  \Delta Z_{n-1}X_{n} X_{n+1}Z_{n+2} +  Z_{n-1}Y_{n} Y_{n+1}Z_{n+2} +  Z_nZ_{n+1} + h  Z_{n-1}Y_nZ_{n+1}\Big)
\end{align} 
is unitarily equivalent to $H_\textrm{XXZ}- h\sum_n Z_n$. To derive this, we conjugate the Hamiltonian by $U_\mathrm{chain}$ built from the basic Clifford transformation $U \in \mathcal{C}_2$ given by:
\begin{align}
X_1 \rightarrow X_1 Z_2 \qquad  &X_2 \rightarrow Z_2  \nonumber \\
Z_1 \rightarrow Z_1\qquad &Z_2 \rightarrow Z_1 Y_2.
\end{align} 
\subsubsection{The folded XXZ chain and pivot Hamiltonians}\label{sec:folded}
As an aside, we note an interesting connection between the model ${H}_{\textrm{XXZ}}'$, the folded XXZ chain \cite{Yang20,Zadnik21,Zadnik21b,Pozsgay21,Gombor21} and the idea of pivot Hamiltonians \cite{Tantivasadakarn2021,Tantivasadakarn2021b}. The folded XXZ chain is an integrable model with many interesting properties, including Hilbert space fragmentation \cite{Sala20,Khemani20}, corresponding to exponentially large degeneracies in the spectrum. 
Pivot Hamiltonians generate SPT entanglers (an example of which is the transformation \eqref{eq:clustermapping}), and have been used to generate webs of dualities between models and to find interesting critical points with enhanced symmetry \cite{Tantivasadakarn2021,Tantivasadakarn2021b}. Indeed, the critical points `half-way between' two models often have the pivot Hamiltonian as an additional U(1) symmetry \cite{Tantivasadakarn2021}.
In this section, we show that using two different Ising Hamiltonians as a pivot, starting with  ${H}_{\textrm{XXZ}}$ we arrive at the model ${H}_{\textrm{XXZ}}'$. The folded XXZ chain is at the half-way point between these two models, and we see that the two Ising Hamiltonians give two additional U(1) symmetries of the folded XXZ chain. While these symmetries were already known, we hope that this connection will lead to new insights.

Let us define the following commuting charges:
\begin{align}
Q_1 &= \frac{1}{2}\sum_{n\in\textrm{sites}} (1-Z_n)\qquad Q_2 =\frac{1}{2}\sum_{n\in\textrm{sites}}(1-Z_nZ_{n+1})\nonumber\\
Q_4 &= -\frac{1}{4}\sum_{n\in\textrm{sites}} (X_{n+1} X_{n+2} + Y_{n+1} Y_{n+2})\left(1+ Z_{n}Z_{n+3}\right).
\end{align}
The Hamiltonian, $H_\mathrm{F}$, of the folded XXZ chain is given by:
\begin{align}
H_{\mathrm{F}} = Q_4 + hQ_1 + \Delta Q_2.
\end{align}
The model $H_{\mathrm{F}}$ also commutes with the operator $Q_2^-=\sum_n (-1)^n Z_n Z_{n+1}$ \cite{Zadnik21}. Now, considering the XXZ chain \eqref{eq:H_XXZ} with an external field, and the transformed chain \eqref{eq:H'XXZ}, we have (up to unimportant constants):
\begin{align}
4 H_{\mathrm{F}} = \underbrace{-\sum_{n \in \textrm{sites}}\Big(  X_{n}X_{n+1}+Y_{n}Y_{n+1} +  \Delta Z_{n}Z_{n+1} + h Z_n\Big)}_{H_{\mathrm{XXZ}}(h)} + \underbrace{\vphantom{\sum_{n \in \textrm{sites}}}U^{\vphantom \dagger}_{\pm} H_{\mathrm{XXZ}}(h) U_{\pm}^\dagger}_{H'_{\mathrm{XXZ}}(h)},\end{align}
where $U_{\pm}$ is a Clifford transformation that can be written as a $U_{\mathrm{chain}}$ with $k=2$ and acts by conjugation as: 
\begin{align}
X_n &\rightarrow  \pm Z_{n-1}X_nZ_{n+1}\nonumber\\ 
Z_n &\rightarrow Z_n. \label{eq:clustermapping2}
\end{align} 
We see that $H_\textrm{F}$ is at the half-way point on a path of models interpolating between the usual XXZ model (with external field), and the model $H'_{\mathrm{XXZ}}$ discussed above.

To make the connection to pivot Hamiltonians, we can write these choices of  $U_\pm$ as $U_\pm(\pi)$ where:
\begin{align}
 U_\pm(\theta) = \rme^{-\rmi \theta H^\pm_{\textrm{pivot}}} \qquad \textrm{for}\qquad H^+_{\textrm{pivot}} =\frac{1}{4} \sum_{n \in \textrm{sites}} (-1)^n Z_nZ_{n+1},\qquad H^-_{\textrm{pivot}} =\frac{1}{4} \sum_{n \in \textrm{sites}} Z_nZ_{n+1}.
\end{align} 
Then define $H^\pm(\theta) = U_\pm(\theta) H_0 U_\pm^\dagger(\theta)$. These are particular examples of the pivot Hamiltonian construction given in Ref.~\cite{Tantivasadakarn2021}. More generally we define $H(\theta)$ in this way for some choice of $H_{\textrm{pivot}}$; where there is a further requirement that $H(2\pi)=H(0)$ and that $H(\pi)$ is an SPT phase for some symmetry group. An example would be the trivial paramagnet $H_0 =-\sum_n X_n$; then $H^+(\pi)$ is the SPT cluster model. In our case we will take $H_0 = H_\mathrm{XXZ}$ which is a critical model; the analogue of the SPT requirement would then be that $H^\pm(\pi)$ is a symmetry-enriched gapless model \cite{Verresen21} for some symmetry group. Analysing the symmetry properties of these models\footnote{For external field $h=0$, there is a $\mathbb{Z}_2\times\mathbb{Z}_2$ symmetry generated by $\prod_n X_n$ and $\prod_n Y_n$. There is also a $\mathbb{Z}_2^T$ antiunitary time-reversal symmetry.} is beyond our scope, we defer it to future work. Understanding this could lead to new perspectives on the folded XXZ model.

One interesting consequence of making this connection is as follows. We can consider a family of models $H^\pm_\alpha = \alpha H^\pm(0) + (1-\alpha) H^\pm(\pi)$. At the half-way point $H^\pm_{1/2}= \frac{1}{2} \left( H^\pm(0) + H^\pm(\pi) \right)$, there is an enhanced symmetry. Indeed, we clearly have $[H^\pm_{1/2},U_\pm(\pi)]=0$ (in our case of interest $U_\pm(\pi)=U_\pm$, we have $U(2\pi) = 1$ and this is a $\mathbb{Z}_2$ symmetry). If $H_0$ is a trivial paramagnet, in many cases this is enhanced to a U(1) symmetry generated by $H^\pm_{\textrm{pivot}}$, i.e., $[H^\pm_{1/2}, H^\pm_{\textrm{pivot}}]=0$. This follows from a general result given in \cite{Tantivasadakarn2021}, based on a well-chosen Clifford transformation (in this context also called a generalised Kramers-Wannier duality or the gauging map in quantum information theory), that reveals a $\mathbb{Z}_n$ symmetry and shows that neutrality of the Hamiltonian under this $\mathbb{Z}_n$ implies neutrality under the full U(1).

Now taking $H_0 = H_\mathrm{XXZ}$, we have that $H^\pm_{1/2} = H_\textrm{F}$. The result given above for trivial $H_0$ no longer applies, but the property that  $[H^\pm_{1/2}, H^\pm_{\textrm{pivot}}]=0$ in this case corresponds to the commutativity of $Q_1, Q_2, Q_2^-$ and $Q_4$. This can be checked straightforwardly in this model, and naturally raises the question of how to extend the argument given in \cite{Tantivasadakarn2021} to further families of Hamiltonian $H_0$, including gapless models.

\subsection{Locality-non-preserving transformations with \texorpdfstring{$k=2$}{k = 2} and the XXZ chain} \label{sec:nonlocal}
We now consider locality-non-preserving transformations $U_\textrm{chain}$, again built from a basic Clifford transformation $U\in\mathcal{C}_2$. By considering all locality-non-preserving $U_{\mathrm{chain}}$ with $k=2$, we show in Section \ref{sec:twositeClifford} below that there are three types of non-local Pauli strings that can appear. Two of these types of string will not preserve locality of $H_{\mathrm{XXZ}}$, while the third
does---this includes the Kramers-Wannier duality given in \eqref{eq:KWtransform}. Using this transformation, we have that the following Hamiltonian is unitarily equivalent\footnote{This is the bulk form of the Hamiltonian, we are not including any boundary terms.} to $H_{\textrm{XXZ}} - h\sum_n Z_n$: 
\begin{align}
{{H}_{\textrm{XXZ}}''} &= -\sum_{n\in\textrm{sites}} \Big(Z_n + \Delta X_{n} X_{n+2}- X_{n-1} Z_n X_{n+1}+ h X_{n} X_{n+1}\Big). \label{eq:H''XXZ}
\end{align}
We conclude that the Kramers-Wannier duality is the only locality-non-preserving transformation (in the class under consideration, and up to on-site basis changes) that takes $H_{\textrm{XXZ}} - h\sum_n Z_n$ to a local Hamiltonian.
Note that ${H}_{\textrm{XXZ}}''$ is related to the anisotropic next-nearest-neighbour Ising (ANNNI) model \cite{Peschel1981}. Using Kramers-Wannier duality to map the XXZ chain to this `$U(1)$-enhanced'   ANNNI model is discussed in \cite{Verresen2019}. The Hamiltonian \eqref{eq:H''XXZ} has also appeared recently in Ref.~\cite{Gombor21} as an example of a medium-range integrable model (the model that appears is in fact the bond-site transformed XYZ chain, rather than XXZ chain; the bond-site transformation corresponds to the Kramers-Wannier transformation used here). There it was derived as part of a classification of `interaction-round-a-face' models with three-site interaction terms.

 We give two other families of locality-non-preserving transformations below. In the Kramers-Wannier case we have strings of the form $X_n \rightarrow Z_nZ_{n+1}Z_{n+2}\cdots$, hence, $X_nX_{n+1} \rightarrow Z_n$. As mentioned, the other locality-non-preserving transformations have different strings, and these do not cancel when we transform operators such as $X_nX_{n+1}$ that appear in $H_{\textrm{XXZ}}$. We can, however, use these locality-non-preserving transformations to map between local integrable Hamiltonians.  For example, consider $H''_{\textrm{XXZ}}$ with $h=0$. Then the transformation given in \eqref{eq:nonlocaltransform2example} transforms this model to:
\begin{align}
{{H}_{\textrm{XXZ}}'''} = -\sum_{n\in\textrm{sites}} \Big(Z_n - X_{n} X_{n+2}+ \Delta X_{n-1} Z_n X_{n+1}\Big). 
\end{align}

\subsection{Classification of \texorpdfstring{$U_\textrm{chain}$}{Uchain} built from a two-site basic Clifford.}\label{sec:twositeClifford}
To prove these results, we now analyse the effect of conjugating \emph{once} by $U_\textrm{chain}$ on single-site Pauli operators. Using the formula given in Section \ref{sec:Cliffordgroup}, we have that $\lvert \mathcal{C}_2 \rvert =11520$, and hence can write down that many different $U_\textrm{chain}$. We will show that these fall into seven classes: three (L1, L2, L3) that are locality-preserving and four (NL1, NL2, NL3, NL4) that are locality-non-preserving. 

Consider any two anti-commuting single-site Pauli operators $P, Q \in \mathcal{P}$, we prove in Section \ref{sec:derivation} that the locality-preserving transformations are of the form:
\begin{enumerate}
\item[(L1)] On-site change of basis: $X_n \rightarrow \pm P_n$, $Z_n \rightarrow \pm Q_n$.
\item[(L2)] On-site change of basis and shift by one site to the left\footnote{Note that there is a corresponding transformation that shifts to the right if we instead conjugate by $U_\textrm{chain}^\dagger$.}: $X_n \rightarrow \pm P_{n-1}$, $Z_n \rightarrow \pm Q_{n-1}$.
\item[(L3)] Decorating transformations \cite{Chen14}
\begin{align}
&X_n \rightarrow \pm S_{n-1}P_n S_{n+1}\nonumber\\
&Z_n \rightarrow \pm S_{n-1}Q_nS_{n+1}, \qquad   \mathrm{where~} S_n := \rmi  P_n Q_n. \label{eq:twosite}
\end{align}
\end{enumerate}
The transformation (L3) could of course be preceded by an on-site change of basis---this will correspond to a $U_\mathrm{chain}$ given by a different basic Clifford transformation in $\mathcal{C}_2$. 
Then, for example, we can also have:
\begin{align}
&X_n \rightarrow \pm S_{n-1}P_n S_{n+1}\nonumber\\
&Y_n \rightarrow \pm S_{n-1}Q_nS_{n+1}, \qquad   \mathrm{where~} S_n := \rmi  P_n Q_n. 
\end{align}
Below, we will take as given that you can always do this initial on-site change of basis.

We now consider locality-non-preserving transformations. These all map at least one Pauli operator to a non-local string of Pauli operators. For a finite system this string will end at the right-most site (this site has index $L$). As we are interested in transforming local Hamiltonians to local Hamiltonians, the precise details of the boundary terms are not so important: they must drop out in local expressions---we use $S'$ and $T$ to represent the Pauli operators that are included in this boundary term. 
We prove in Appendix \ref{app:nonlocal} that the locality-non-preserving transformations are of the forms (NL1)-(NL4) below. (Note that there exist transformations $U_\mathrm{chain}$ for all choices of anti-commuting $P, Q \in \mathcal{P}$; the operator appearing in the string, $S$, is then fixed.)
\begin{enumerate}
\item[(NL1)] Transformations leading to a string $\prod_{j} S_{n+j}$:
\begin{align}
&X_n \rightarrow  (-1)^{an+b}\phantom{S''_{n-1}} S_nS_{n+1}S_{n+2}S_{n+3}\cdots S_{L-1} T_{L} \nonumber\\
&Z_n \rightarrow (-1)^{an + c} P_{n-1}Q_n S_{n+1}S_{n+2}S_{n+3}\cdots S_{L-1} T_{L},  \label{eq:nonlocaltransform1}
\end{align}
where $S_n := \pm\rmi  P_n Q_n$ and $a,b,c\in\{0,1\}$.
\item[(NL2)] Transformations leading to a string $\prod_{j} S_{n+2j}$: 
\begin{align} 
&X_n \rightarrow  (-1)^{x(n)} P_{n-1}S_n S_{n+2}S_{n+4}S_{n+6}\cdots S'_{L-1}  T_{L}  \nonumber\\
&Z_n \rightarrow (-1)^{z(n)} Q_{n-1}S_n S_{n+2}S_{n+4}S_{n+6}\cdots {S}'_{L-1}  T_{L}, \label{eq:nonlocaltransform2}
\end{align}
where $ S_n:= \pm \rmi  P_n Q_n$ and the oscillatory terms depend on $x(n)$ and $z(n)$ that are discussed in Appendix \ref{app:nonlocal}.
\item[(NL3)] Transformations leading to a string given by $\prod_{j} S_{n+3j}S_{n+3j+1}$:  
\begin{align}
X_n &\rightarrow (-1)^{x(n)} P_{n-1}S_n S_{n+1}S_{n+3}S_{n+4}S_{n+6}S_{n+7}\cdots {S}'_{L-1}T_{L}  \nonumber\\
Z_n &\rightarrow (-1)^{z(n)} Q_{n-1}S_n S_{n+1}S_{n+3}S_{n+4}S_{n+6}S_{n+7}\cdots {S}'_{L-1} T_{L}, \label{eq:nonlocaltransform3}
\end{align}
where $S_n := \pm\rmi  P_n Q_n$ and the oscillatory terms depend on $x(n)$ and $z(n)$ that are discussed in Appendix \ref{app:nonlocal}.
\item[(NL4)] On-site change of basis and shift by one site to the left, with boundary term and oscillatory factor:  
\begin{align}
&X_n \rightarrow (-1)^{an+b}  P_{n-1}  T_{L}  \nonumber\\
&Z_n \rightarrow (-1)^{an+c} Q_{n-1}   T_{L} ,
\label{eq:nonlocaltransform4}
\end{align}
where $a,b,c\in\{0,1\}$.\end{enumerate}
As mentioned, $S'$ and $T$ are the boundary terms---these are unimportant for transforming local Hamiltonians to local Hamiltonians. The operator  $S'$ is either equal to $S$ or to $\mathbb{I}$ depending on whether the site $L-1$ would appear in the string\footnote{For example, for the string of the form $\prod_{j} S_{n+2j}$, then $S'=S$ if $(L-1-n)$ is even, and $S'=\mathbb{I}$ if $(L-1-n)$ is odd. One can see that $S'$ must fit the string pattern on site $L-1$ since if we increase the length of the chain, then $U_\textrm{chain}\rightarrow U_{L,L+1} U_\textrm{chain}$, and $U_{L,L+1}$ does not act on this site.}; $T$ is not fixed (and in general depends on $S'$, $L$ and details of the particular Clifford transformation---for example, we show in Appendix \ref{app:nonlocal} that one can write down a Clifford transformation leading to \eqref{eq:nonlocaltransform1} with any choice of $T\in \mathcal{P}$). Note also that these are bulk transformations, so the transformation will, in general, act differently on sites at the boundary. 
\subsection{Derivation of this classification}\label{sec:derivation}
To prove the claim about locality-preserving transformations, let us define\footnote{This is a partial definition: to fully define the Clifford transformation we also need the image of $X_{n+1}$ and of $Z_{n+1}$.} a two-site Clifford transformation $U_{n,n+1}$ acting non-trivially on sites $n$ and $n+1$ by:
\begin{align}
X_n \rightarrow s_X & L_{n\phantom{+1}}^{(X)} R_{n+1}^{(X)},\quad Z_n \rightarrow s_Z L_{n\phantom{+1}}^{(Z)} R_{n+1}^{(Z)} \qquad \Big(\Rightarrow\quad Y_n \rightarrow s_Xs_Z \rmi L_{n\phantom{+1}}^{(X)}L_{n\phantom{+1}}^{(Z)}  R_{n+1}^{(X)}R_{n+1}^{(Z)}\Big) . \label{eq:twositeclifford}
\end{align}
$L$ and $R$ are single-site Pauli operators on the left and right sites\footnote{For example, for the transformation in \eqref{eq:cluster2site} we have $L^{(X)}=X$, $R^{(X)}=Z$, $L^{(Z)}= Z$ and $R^{(Z)}=\mathbb{I}$.} and $s_X,s_Z\in\{\pm1\}$. It is enough to consider how the transformation $U_\textrm{chain}$ acts left to right, since that is the direction where $U_\textrm{chain}$ has a non-local causal structure. There are three cases to consider: \begin{enumerate}
    \item  First, if $R^{(X)}=R^{(Z)} = \mathbb{I}$ then this is a single-site Clifford transformation and we simply have an on-site change of basis: the transformations of the form (L1). 
    \item The second case is where $R^{(X)} \neq R^{(Z)}$ and both are non-trivial, implying $R^{(Z)}\neq R^{(X)}R^{(Z)}\neq R^{(X)}$. Hence, all three single-site Pauli operators appear in the $R$-terms on site $n+1$ in \eqref{eq:twositeclifford}. This means that $X_n, Y_n$ and $Z_n$ will keep growing into non-local strings of operators as we continue to apply the transformations making up $U_\textrm{chain}$. The only exception to this is when in addition $L^{(X)}=L^{(Z)}=\mathbb{I}$---this does keep bulk operators local, corresponding to an on-site change of basis and a shift by one site to the left\footnote{If we have a periodic system then this does shift every site one site to the left, since this will take $P_1\rightarrow P'_L$. The corresponding MPU is an on-site change of basis followed by the left-moving shift operator \cite{Cirac21}.}. Hence the locality-preserving transformations in this case give the transformations of the form (L2). 
    \item The final case is where one of $R^{(X)}$, $R^{(Z)}$ and $R^{(X)}R^{(Z)}$ is equal to identity (and then the other two must be equal). Let us consider a non-trivial $R^{(X)} =R^{(Z)} = R$. If $R = Y$ then we have a locality-preserving  $U_{\mathrm{chain}}$. This follows since $R^{(Y)}=\mathbb{I}$ and so
    \begin{align}
    U_{n+1,n+2}^{\vphantom\dagger}\left(U_{n,n+1}^{\vphantom\dagger}X_n^{\vphantom\dagger}U^\dagger_{n,n+1}\right)U_{n+1,n+2}^\dagger &= s_XU_{n+1,n+2}^{\vphantom\dagger}\left(L^{(X)}_{n\phantom{+1}}Y_{n+1}^{\vphantom\dagger}\right)U_{n+1,n+2}^\dagger\nonumber\\ 
    &= s_Z L^{(X)}_{n\phantom{+1}} \left(\rmi L_{n+1}^{(X)}L_{n+1}^{(Z)}\right).\label{eq:R=Y}
    \end{align}
    Further two-site Clifford transformations making up $U_\textrm{chain}$ act on sites to the right of $n+1$ so we see it transforms $X_n$ to a local operator. 
    An analogous calculation holds for transforming $Z_n$, and so for $R=Y$ we have a locality-preserving transformation.  If $R \neq Y$ then we will have non-local strings by iterating the transformation. To see this explicitly:
    \begin{align}
    U_{n+1,n+2}^{\vphantom\dagger}\left(U_{n,n+1}^{\vphantom\dagger}X_n^{\vphantom\dagger}U^\dagger_{n,n+1}\right)U_{n+1,n+2}^\dagger&= s_XU_{n+1,n+2}^{\vphantom\dagger}\left(L^{(X)}_{n\phantom{+1}}R_{n+1}^{\vphantom\dagger}\right)U_{n+1,n+2}^\dagger \nonumber\\&= s_Xs_RL^{(X)}_{n\phantom{+1}} L^{(R)}_{n+1}R_{n+2},
    \end{align}
    where further two-site Clifford transformations making up $U_\textrm{chain}$ will continue to act non-trivially. We can make an analogous argument for the case that $R^{(X)}=\mathbb{I}$ and also for the case that $R^{(Z)}=\mathbb{I}$. For example, if $R^{(Y)}=R^{(Z)}=X$ we have locality-preserving transformations, while if $R^{(Y)}=R^{(Z)}=Y$ we will not.
    
    In \eqref{eq:R=Y} we saw that if $R=Y$, then the transformation is locality-preserving. We now consider how $U_\mathrm{chain}$ acts in this case. Firstly, $R=Y$ means that in \eqref{eq:twositeclifford} we must have $L^{(X)}\neq L^{(Z)}$. We can then consider how $U_{n,n+1}$ in \eqref{eq:twositeclifford} acts on $X_{n+1}$ and $Z_{n+1}$. One can show that\footnote{To see this, consider $X_{n+1}\rightarrow P_n Q_{n+1}$. If $P \in \{L^{(X)},L^{(Z)}\}$, we are led to an inconsistency since conjugation by $U_{n,n+1}$ preserves the commutators $[X_n,X_{n+1}]=[Z_n,X_{n+1}]=0$.}, up to signs, the images of $X_{n+1}$ and $Z_{n+1}$ are any two distinct elements of $\{Y_{n+1}, \left(\rmi L^{(X)}_nL^{(Z)}_n\right)X_{n+1},\left(\rmi L^{(X)}_nL^{(Z)}_n\right)Z_{n+1}\}.$
    Note that the choice of which operator goes to which operator is simply the choice of initial basis for our transformation $U_{\mathrm{chain}}$. Let us define the product $S:=\rmi L^{(X)}L^{(Z)}$. We see that, without loss of generality, conjugating by $U_{\mathrm{chain}}$ acts as:
\begin{align}
&X_n \rightarrow \pm S_{n-1}L^{(X)}_n S_{n+1}\nonumber\\
&Z_n \rightarrow \pm S_{n-1}L^{(Z)}_nS_{n+1}.
\end{align}
    In this way we reach the locality-preserving transformations (L3).
\end{enumerate}

To prove the claim about locality-non-preserving transformations, we simply have to consider all the cases we excluded in the above discussion. In particular, we need to consider the cases where $R^{(X)} = R^{(Z)}\neq Y$ and where $R^{(X)} \neq R^{(Z)}$ and both are non-trivial. These different cases are analysed in Appendix \ref{app:nonlocal}, here we give examples of a basic Clifford transformation that generates each type. 

\begin{enumerate}
    \item[(NL1)] Up to an on-site basis change, this is the Kramers-Wannier duality with basic Clifford transformation \eqref{eq:KWtransform}.
\item[(NL2)] Putting $P=X$, $Q=Y$ and $S=Z$, we have the basic Clifford transformation:
\begin{align} X_n\rightarrow -Z_nX_{n+1} \qquad &X_{n+1}\rightarrow X_{n}Z_{n+1}\nonumber\\
Z_n\rightarrow Z_n Y_{n+1} \qquad &Z_{n+1}\rightarrow Y_nZ_{n+1}. \label{eq:nonlocaltransform2example}
\end{align} 
This example has $x(n)=z(n)=0$; so the string is not oscillatory.
\item[(NL3)] Putting $P=Z$, $Q=Y$ and $S=X$, we have the basic Clifford transformation:
\begin{align} X_n\rightarrow X_nZ_{n+1} \qquad &X_{n+1}\rightarrow Z_{n}X_{n+1}\nonumber\\
Z_n\rightarrow X_n Y_{n+1} \qquad &Z_{n+1}\rightarrow Y_nX_{n+1}. \end{align} 
This example has $x(n)=z(n)=0$; so the string is not oscillatory.
\item[(NL4)]Putting $P=X$ and $Q=Z$, we have
\begin{align} X_n\rightarrow -Y_nX_{n+1} \qquad &X_{n+1}\rightarrow X_{n}Y_{n+1}\nonumber\\
Z_n\rightarrow Y_nZ_{n+1} \qquad &Z_{n+1}\rightarrow Z_nY_{n+1}.
\end{align} 
Since $Y_n\rightarrow -Y_{n+1}$ we have an oscillatory term $\pm (-1)^{n}$, and, moreover, $T=Y$.
\end{enumerate} 

We note that the results for locality-preserving transformations given above are consistent with the theory of Clifford QCAs \cite{Schumacher2004,Schlingemann2008}, as they should be. However, to prove our results using that theory, one would first need to show that the operators under consideration have images on up to three neighbouring sites. 
\begin{table*}
\begin{center}\begin{tabular}{|c|ccc|c|}\hline  & &Basic Clifford & ($U_{12}$ or $U_{123}$)& Clifford group transformation ($U_{\mathrm{chain}}$) \\
\hline $\tilde{U}^{\vphantom \dagger}_1$  & $X_1 \rightarrow X_1 Z_2$ &  $X_2 \rightarrow Z_1 Y_2$ & & $X_n \rightarrow Z_{n-1}Y_nZ_{n+1}$ \\
&$Z_1 \rightarrow Z_1$& $Z_2 \rightarrow Z_2$ && $Z_n \rightarrow Z_n$ \\
\hline $\tilde{U}^{\vphantom \dagger}_2$  &   $X_1 \rightarrow Y_1Z_2Z_3 $&  $X_2 \rightarrow Z_1 Y_2$ & $X_3 \rightarrow Z_1Y_3$&$X_n \rightarrow  -Z_{n-2}Z_{n-1}Y_nZ_{n+1}Z_{n+2}$\\
 & $Z_1 \rightarrow Z_1 $&  $Z_2 \rightarrow Z_2$ & $Z_3 \rightarrow Z_3$ & $Z_n \rightarrow  Z_n$\\
\hline $\tilde{U}^{\vphantom \dagger}_3$  &  $X_1 \rightarrow X_1Z_2Z_3 $&  $X_2 \rightarrow Z_1X_2Z_3$ & $X_3 \rightarrow Z_1Z_2X_3$ &$X_n \rightarrow Z_{n-2}X_nZ_{n+2}$  \\
 & $Z_1 \rightarrow Z_1 $&  $Z_2 \rightarrow Z_2$ & $Z_3 \rightarrow Z_3$& $Z_n \rightarrow Z_n$\\
\hline $\tilde{U}^{\vphantom \dagger}_4$ &  $X_1 \rightarrow X_1X_2X_3 $&  $X_2 \rightarrow Z_1Y_2X_3$ & $X_3 \rightarrow Z_1X_2Y_3$  & $X_n \rightarrow -Z_{n-2}Y_{n-1}Y_nY_{n+1}Z_{n+2}$\\ & $Z_1 \rightarrow Z_1 $&  $Z_2 \rightarrow X_2$ & $Z_3 \rightarrow X_3$& $Z_n  \rightarrow -Z_{n-1}X_nZ_{n+1}$\\
\hline
\end{tabular}\vspace{0.15cm} \caption{Transformations $U_\textrm{chain}$ of the type shown in Figure \ref{fig:uchain} that take $H_\textrm{XXZ}$ to $H_i$ by $H_i=\tilde{U}_iH_\textrm{XXZ}\tilde{U}_i^\dagger$.}\label{table:transformationstable}
\end{center}
\end{table*}

\section{Generalisations---further families related to the XXZ model}\label{sec:generalisations}
In this section we give a classification of all translation invariant Clifford transformations on the spin chain that map $X_n$ and $Z_n$ to Pauli strings on at most five neighbouring sites (i.e., each image is of the form $P^{(1)}_{m-2}P^{(2)}_{m-1}P^{(3)}_{m\vphantom{+1}}P^{(4)}_{m+1}P^{(5)}_{m+2}$). This is the same mathematical problem as classifying Clifford QCAs where the images of $X_n$ and $Z_n$ are non-trivial on at most five neighbouring sites\footnote{\label{footnote:QCA}Note that this implies that the images of $X_n$ and $Z_n$ are restricted to a single set of five neighbouring sites. This follows from the general result for Clifford QCAs that the images of $X_n$ and $Z_n$ must be symmetric under reflection about a site, and that this is the same site for both images \cite{Schlingemann2008} (we give a derivation for the case under consideration in Appendix \ref{app:constraints}). This means that the image of $Y_n$ is also symmetric about the same site, with support restricted to the support of the images of $X_n$ and $Z_n$.}. We furthermore show that all of these transformations are of the form $U_\textrm{chain}$ for some basic Clifford $U \in \mathcal{C}_2$ or $U\in \mathcal{C}_3$. We then use these results to write down a form for all Hamiltonians that are equivalent to $H_\textrm{XXZ}$ under translation invariant Clifford group transformations that take $X_n$ and $Z_n$ to local operators on at most five sites.

In the previous section we started by considering basic Clifford transformations, and showed that certain families of $U_\textrm{chain}$ were locality-preserving. An alternative approach is to consider locality-preserving transformations directly.  As discussed in Section \ref{sec:Cliffordgroup}, unitary transformations preserve commutation relations. If our unitary is a translation invariant Clifford group transformation, this gives a map $X_n \rightarrow P $ and $Z_n \rightarrow Q $, with $P$ and $Q$ some Pauli strings. Commutators such as  
$[X_n,X_{n+m}]=0 \Leftrightarrow [P,P(m)]=0$ give us non-trivial constraints on $P$ and $Q$ (recall that $P(m)$ is defined in Section \ref{sec:Cliffordgroup} as a translation of $P$ by $m$ sites) .
In Appendix \ref{app:constraints} we analyse these constraints in the case that $P$ and $Q$ are Pauli strings on up to five sites. 
The outcome is that if we map any single-site Pauli operator to a two- or four-site operator, then it must map the other on-site Pauli operators to non-local strings (this follows from the theory of Clifford QCAs---see Footnote \ref{footnote:QCA}). We also prove the results for locality-preserving transformations given in the next subsection. 
\subsection{Locality-preserving transformations with images on at most five neighbouring sites:}\label{sec:generalisationsresults}
If we map to a three- or five-site operator we show below that we have the transformations that we already found using basic Clifford transformations in $\mathcal{C}_2$, along with the following additional transformations:
\begin{enumerate}
\item[(L4)] \textit{Decorating transformation:} \begin{align}
&X_n \rightarrow\pm A_{n-2}A_{n-1}C_n A_{n+1}A_{n+2}\nonumber\\
&Z_n \rightarrow\pm A_{n-2}A_{n-1}C'_n A_{n+1}A_{n+2},
\end{align}
with $A,C,C' \in \mathcal{P} \setminus \{\mathbb{I}\}$, $C\neq C'$ and $A=\rmi CC'$.
 \item[(L5)]\textit{Decorating transformation:}  \begin{align}
&X_n \rightarrow\pm A_{n-2}C_nA_{n+2}\nonumber\\
&Z_n \rightarrow\pm A_{n-2}C'_n A_{n+2},
\end{align}
with $A,C,C' \in \mathcal{P} \setminus \{\mathbb{I}\}$, $C\neq C' $ and $A=\rmi CC'$.
\item[(L6)] \textit{Decorating transformation:} 
\begin{align}
&X_n \rightarrow \pm A_{n-2}B_{n-1}A_n B_{n+1}A_{n+2}\nonumber\\
&Z_n \rightarrow \pm A_{n-2}C_{n-1}C_n C_{n+1}A_{n+2} \end{align}
with $A, B, C \in \mathcal{P} \setminus \{\mathbb{I}\}$, $A\neq B$ and $C=\rmi AB$.
\end{enumerate}
As above, we can always precede these transformations by an on-site change of basis; equivalent to taking a different $U_\mathrm{chain}$. Furthermore, one can show that (L4)-(L6) can all be derived using $U_\mathrm{chain}$ built from $U \in \mathcal{C}_3$. This is demonstrated\footnote{Note that a change of basis after the transformation is needed to reach the general transformations given above. To find the corresponding basic Clifford transformation, take the transformation from Table \ref{table:transformationstable} and then transform the Pauli operators on the first site on the right-hand-side of the transformation.} in Table \ref{table:transformationstable}. Finally, note that by considering only the constraints from translation invariance and the commutation relations, we can include an arbitrary translation (or shift operator). We have fixed this translation so that the images are symmetric about site $n$---transformations $U_{\mathrm{chain}}$ do not include such an arbitrary translation. 

\begin{figure}[t]
\centering
\includegraphics{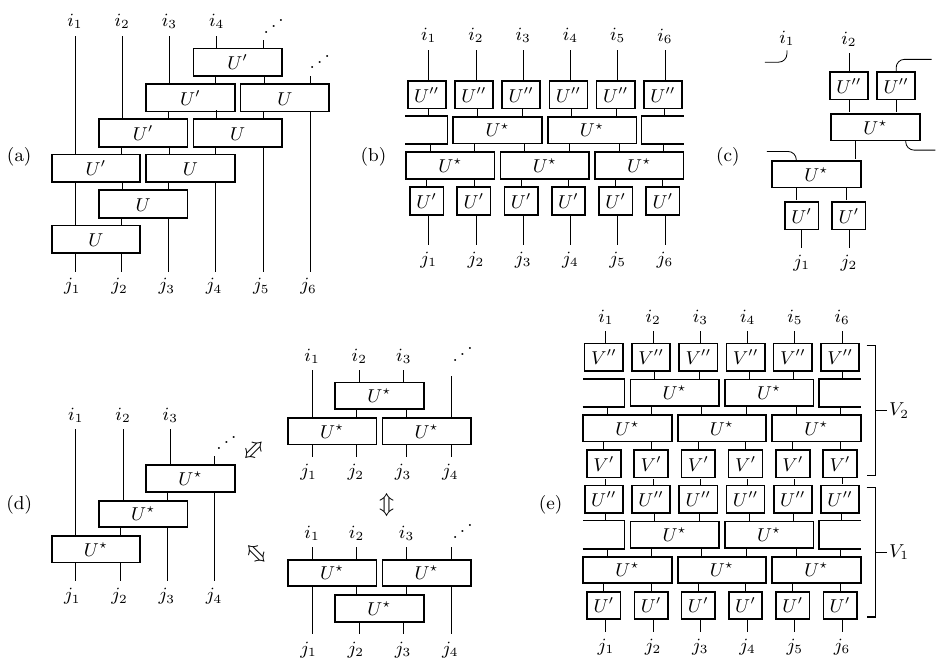}
 \caption{(a) Applying  $U_\textrm{chain}$ followed by $U'_\textrm{chain}$ generates further transformations. (b) A depth-four local unitary quantum circuit; Figure \ref{fig:uchain}(a) is of this form for $U^\star$ given in \eqref{eq:U12}. (c) An MPU tensor corresponding to (b) (note that we block neighbouring spins, so the physical dimension is four). (d) Reordering $U^\star_\mathrm{chain}$ built from $U^\star$ that commute on different pairs of sites. (e) A depth-eight circuit that has the same effect as transforming with $\tilde{U}_4$.  }
    \label{fig:uchaincommuting}

\end{figure}
\subsection{Families of models equivalent to the XXZ chain under locality-preserving transformations:}\label{sec:summary}
We can summarise some of the above results as follows. We have that the spin chains:
\begin{align}
{H_1} &= -\sum_{n\in\textrm{sites}}\Big( J_1 Z_{n-1}X_{n} X_{n+1}Z_{n+2} + J_2 Z_{n-1}Y_{n} Y_{n+1}Z_{n+2} + J_3 Z_nZ_{n+1}\Big),\\
{H_2}& = -\sum_{n\in\textrm{sites}} \Big( J_1 Z_{n-2}X_{n} X_{n+1}Z_{n+3} +  J_2 Z_{n-2}Y_{n} Y_{n+1}Z_{n+3} + J_3 Z_nZ_{n+1}\Big), \\
{H_3} &= -\sum_{n\in\textrm{sites}} \Big( J_1 Z_{n-2}Z_{n-1}X_{n} X_{n+1}Z_{n+2}Z_{n+3} + J_2 Z_{n-2}Z_{n-1}Y_{n} Y_{n+1}Z_{n+2}Z_{n+3} + J_3 Z_nZ_{n+1}\Big),
\\
{H_4} &= -\sum_{n\in\textrm{sites}} \Big( J_1 Z_{n-2}X_{n-1}X_{n+2}Z_{n+3} + J_2 Z_{n-2}Y_{n-1}Y_{n}Y_{n+1}Y_{n+2}Z_{n+3} + J_3 Z_{n-1}Y_nY_{n+1}Z_{n+2}\Big),
\end{align} 
where two of the $J_i$ are equal to 1 and the third is equal to $\Delta$, are equivalent to the XXZ chain under Clifford group transformations. Furthermore, these are all unitarily related to $H_\textrm{XXZ}$ by a transformation of the form given in Figure~\ref{fig:uchain}. Moreover, up to an on-site change of basis, this is a \emph{complete list} of Hamiltonians that are equivalent to $H_\textrm{XXZ}$ under translation invariant Clifford group transformations that take $X_n$ and $Z_n$ to local operators on at most five sites---in Table \ref{table:transformationstable} we give the relevant transformations for $J_3 = \Delta$. For the cases with $J_1$ or $J_2=\Delta$, we do an on-site change of basis to 
 $H_\textrm{XXZ}$ and then use a transformation from Table \ref{table:transformationstable}.

It is simple to generate further models by applying a sequence of such transformations (including single-site unitaries). This is illustrated in Figure \ref{fig:uchaincommuting}(a) for the case of two transformations $U'_{\mathrm{chain}}U_{\mathrm{chain}}$. The $\tilde U_i$ defined in Table \ref{table:transformationstable} are such that $H_i = \tilde U_i H_\textrm{XXZ} \tilde U_i^\dagger$. Then $H_0=\tilde U_4\tilde U_2 H_\textrm{XXZ}\tilde U_2^\dagger \tilde U_4^\dagger$ is the Hamiltonian given in \eqref{eq:H_0}. In Appendix \ref{app:H0} we use this unitary transformation to understand the nearby phases to $H_0$ and to show that it is at a phase transition between SPT phases.

\subsection{Commuting Clifford transformations and the MPU form} \label{sec:commuting}

Notice that the two-site Clifford $U^{\star}_{12}$ 
\begin{align}
X_1 \rightarrow X_1Y_2  \qquad & X_2\rightarrow \pm  Y_1 X_2 \nonumber\\
Z_1 \rightarrow  Z_1 Y_2 \qquad & Z_2 \rightarrow \pm Y_1 Z_2, \label{eq:U12}
\end{align}
satisfies $[U^{\star}_{n,n+1},U^{\star}_{n+1,n+2}]=0$ and the corresponding transformation $U^\star_\textrm{chain}$ takes $X_n \rightarrow \pm Y_{n-1}X_nY_{n+1}$ and $Z_n \rightarrow \pm Y_{n-1}Z_nY_{n+1}$. Reordering the $U^\star$ as in Figure \ref{fig:uchaincommuting}(d), we then have a depth-two local unitary quantum circuit for $U^\star_\mathrm{chain}$---illustrated in Figure~\ref{fig:uchaincommuting}(b) (on setting $U'=U''=\Id$). By using on-site basis transformations beforehand and afterwards we can then do any of the transformations (L3). This means that for any decorating transformation of the form (L3) we have a depth-four local unitary quantum circuit, as shown in Figure~\ref{fig:uchaincommuting}(b). (Note that translation invariance is retained due to the commutativity of the $U^\star_{n,n+1}$.) This tells us in particular that the transformation $\tilde U_1$ appearing in Table \ref{table:transformationstable} can be written in this way.

We can similarly find local unitary circuits of depth at most eight for the remaining transformations of Table \ref{table:transformationstable}. One can check that the basic Clifford transformations that appear in $\tilde U_2$ and $\tilde U_3$ commute when applied to different sites. This means that, by analogy with Figure \ref{fig:uchaincommuting}(d), $\tilde U_2$ and $\tilde U_3$ can be written as depth-three local unitary quantum circuits. The basic Clifford transformation appearing in $\tilde U_4$ does not commute when applied to different sites, however, notice that applying two consecutive two-site decorating transformations gives the same effect. In particular, let $V_1$ take $X_n \rightarrow Z_{n-1}Y_nZ_{n+1}$, $Y_n \rightarrow Z_{n-1}X_nZ_{n+1}$ and $V_2$ take $X_n \rightarrow Z_{n-1}X_nZ_{n+1}$, $Z_n \rightarrow Z_{n-1}Y_nZ_{n+1}$. Then conjugation by $\tilde U_4$ is the same as conjugation by $V_2V_1$. This means that $\tilde U_4$ can be written as a local unitary circuit of depth at most eight---this is illustrated in Figure \ref{fig:uchaincommuting}(e), where $U', U''$ and $V',V''$ are the appropriate basis changes for the decorating transformations $V_1$ and $V_2$ respectively. 

This leads to the following conclusion: any $U_\textrm{chain}$ that takes each $X_n$ and each $Z_n$ to a local operator on at most five sites can be written as a local unitary quantum circuit of depth at most eight. (We include on-site changes of basis before and after applying transformations in Table \ref{table:transformationstable}. For $\tilde{U}_4$ we can change $U'$ and $V''$ to give the correct basis.) Moreover, in this form one can see how to explicitly make this transformation periodic. Then, as in going from Figure \ref{fig:uchaincommuting}(b) to Figure \ref{fig:uchaincommuting}(c), one can find an MPU tensor. Having such an MPU tensor is helpful in understanding how the integrable structure of the XXZ model transforms, as we will see in the next section.

\section{Integrable models} \label{sec:integrable}

Thus far, we have considered translation invariant Clifford transformations and used them to construct families of Hamiltonians that are unitarily equivalent to $H_{\textrm{XXZ}}$. The XXZ model, defined in \eqref{eq:H_XXZ}, is an integrable model, and this integrability is related to some underlying structure. It is intuitive that this integrable structure will be roughly unchanged under Clifford transformations, or indeed other unitary transformations. It is the purpose of this section to begin to address in what sense the transformed XXZ chains constructed above share this structure, and to see how this work fits more generally into the classification of integrable models. We will not comment on the precise meaning of the term quantum integrable model, rather we refer the interested reader to the discussion in \cite{Caux}. Pragmatically, if one knows something useful about the states of some Hamiltonian, the ideas introduced above can be used to understand some other Hamiltonian. Two classes about which we can say a lot are free models and models solvable by Bethe Ansatz, such as the XXZ model. Free models are those where there is no quasiparticle scattering, while Bethe Ansatz solvable models have the property that many-quasiparticle scattering can be factorised into two-body scattering. 

We will first analyse how conjugation by a general MPU affects the integrable structure of a Bethe Ansatz solvable model. Then we will explain what this means for the Clifford transformations that we consider. Finally, we will discuss free models. 

\subsection{Models solvable by algebraic Bethe Ansatz}\label{sec:Bethe}
\subsubsection{Generalities}
\begin{figure}[t]\scalebox{0.8}{
\includegraphics{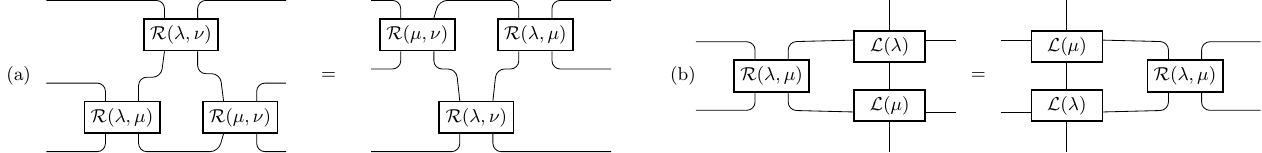}}
\caption{(a) Yang-Baxter equation for $\mathcal{R}$-matrices. (b) Yang-Baxter algebra for $\mathcal{L}$-operators. }
\label{fig:YangBaxter}
\end{figure}

The algebraic Bethe Ansatz is a well developed method that allows us, by solving a system of `Bethe equations', to find the eigenstates and eigenvalues of an integrable Hamiltonian. It is a further non-trivial step to derive correlation functions in these eigenstates, however, there are many known results \cite{Boguliubov1986,Korepin1997,Baxter2016,Franchini,Murg}. We point out that certain results apply at finite system sizes, while others apply only on taking the thermodynamic limit. 
The central object in this method is the $\mathcal{R}$-matrix. This is an operator that solves the Yang-Baxter equation, depicted graphically\footnote{The $\mathcal{R}$-matrix acts on virtual Hilbert spaces $\mathcal{V}_i\otimes \mathcal{V}_j \rightarrow \mathcal{V}_i\otimes \mathcal{V}_j$, and depends on two `spectral parameters' or `rapidities' $\lambda$ and $\mu$: i.e., it is a tensor of the form $\mathcal{R}(\lambda,\mu)_{\alpha_i,\beta_j}^{\gamma_i,\delta_j}$, where each index ranges from 1 to $\chi$ (the bond dimension). These indices are always contracted in calculating physical quantities. We prefer the graphical notation, and refer the reader to \cite{Murg} for a detailed exposition of this.  } in Figure \ref{fig:YangBaxter}(a). 
Using such a solution to the Yang-Baxter equation, one can generate a set of local conserved charges, and one of these is the Hamiltonian of the corresponding \emph{fundamental} integrable model. There is no prescription to go from an integrable Hamiltonian (which may or may not be fundamental) to the $\mathcal{R}$-matrix \cite{Franchini}, although there are interesting tests for integrability of spin chain Hamiltonians \cite{Grabowski,Gombor21}. The XXZ spin chain is a Hamiltonian that is the fundamental model derived from the XXZ $\mathcal{R}$-matrix and can be analysed using the algebraic Bethe Ansatz.

\subsubsection{The fixed-point of an MPU}\label{sec:MPU}
\begin{figure}[t]
\scalebox{0.75}{
\includegraphics{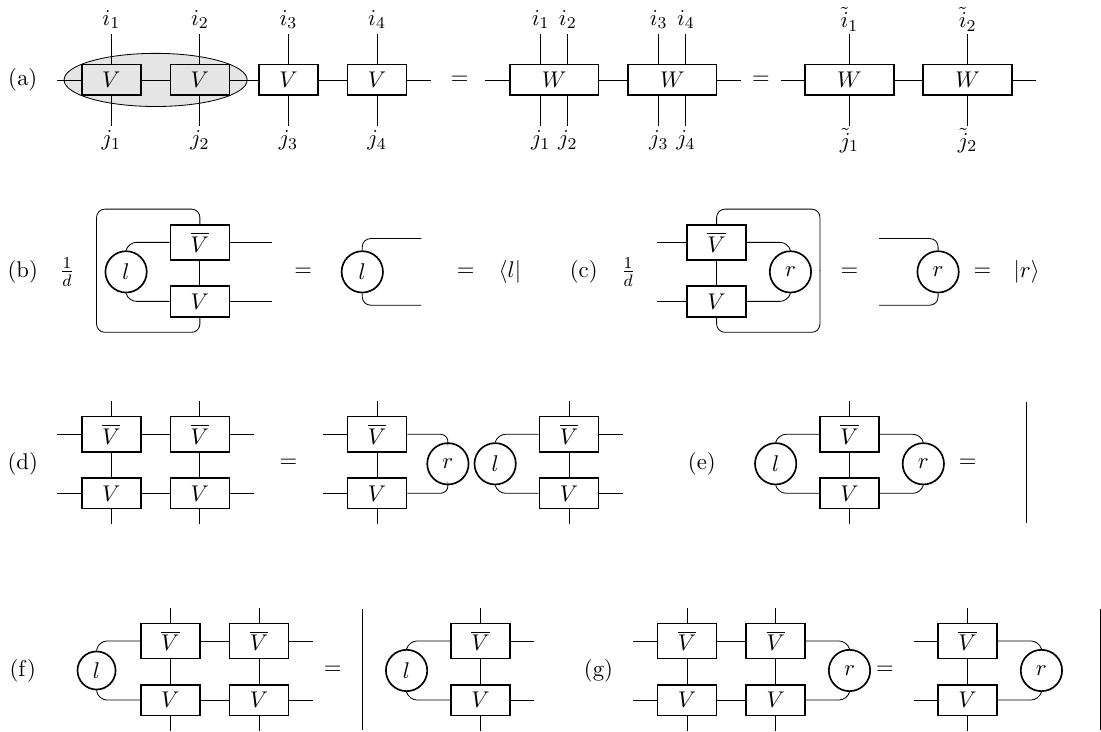}}
\caption{(a) Blocking an MPO; the two shaded tensors $V$ are combined (by contraction of a virtual index) to give a single tensor $W$. (b) and (c) define the vectors $\langle l \rvert $ and  $\lvert r \rangle$ as left- and right- eigenvectors of a matrix formed from the tensors $V$ and $\overline{V}$; $d$ is the dimension of the physical index. (d) The first MPU fixed-point equation.  (e) The second MPU fixed-point equation. (f) and (g) are the `pulling through conditions', a consequence of the fixed-point equations.} 
\label{fig:MPUfixedpoint}
\end{figure}

Before proceeding, we briefly review some further technical results about MPUs that are needed in the following discussion. These results and corresponding proofs are found in Refs.~\cite{Cirac2017,Csahinouglu2018}. All of the results are given diagrammatically in Figure~\ref{fig:MPUfixedpoint}.

We can combine tensors by blocking physical indices, this simply involves viewing multiple physical sites as one physical site with a larger on-site dimension. This is illustrated in Figure~\ref{fig:MPUfixedpoint}(a); if $i_1$ and $i_2$ correspond to a two-dimensional spin-1/2 site, then $\tilde{i}_1$ will be a four-dimensional index corresponding to the tensor product. 
An important result is that if we have an MPU with bond dimension (dimension of the virtual index) equal to $\chi$, then by blocking at most $\chi^4$ sites, we have a fixed-point (or `simple') form for the MPU \cite{Cirac2017,Csahinouglu2018}. 

Now, let us suppose that we have blocked sites such that the MPU is in the fixed-point form. From the corresponding blocked tensors $V$ and $\overline{V}$ we can form a `transfer matrix'. This transfer matrix has just one non-zero eigenvalue, and this is equal to one \cite{Cirac2017}. Then we have two vectors $\langle l \rvert $ and  $\lvert r \rangle$; the unique left- and right-eigenvectors\footnote{We have $\langle l \rvert =\sum_{n=1}^\chi \langle n,n \rvert$ if we gauge-fix the tensor $V$ such that it is in a canonical form \cite{Cirac2017}, but we will not need the explicit form of these vectors for our purposes.} with eigenvalue one---see Figures~\ref{fig:MPUfixedpoint}(c) and (d). The blocked tensor $V$ satisfies two fixed-point equations, given in Figures~\ref{fig:MPUfixedpoint}(d) and (e). These equations imply two further equations called pulling through conditions, given in Figures~\ref{fig:MPUfixedpoint}(f) and (g).

\subsubsection{Integrable models transformed by an MPU}

\begin{figure}[t]
\scalebox{0.75}{
\includegraphics{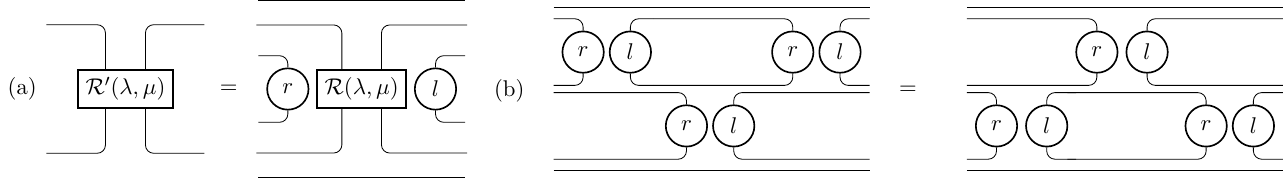}
}
 \caption{(a) The $\mathcal{R'}$-matrix in the enlarged virtual space (the virtual space of the original $\mathcal{R}$-matrix and an additional virtual space). (b) The Yang-Baxter equation in the additional virtual space. This identity holds for any vectors $\lvert r \rangle$ and $\langle l \rvert $, and implies that $\mathcal{R}'(\lambda,\mu)$ satisfies the Yang-Baxter equation if $\mathcal{R}(\lambda,\mu)$ does. Note that this is not a regular solution of the Yang-Baxter equation.} 
 \label{fig:R'matrix}
\end{figure}

Recall that in Section \ref{sec:commuting} we showed that for a given locality-preserving $U_{\mathrm{chain}}$ given in Table \ref{table:transformationstable}, we can find an MPU tensor $V$ such that conjugating by the MPU $M^{i_1,\dots,i_L}_{j_1,\dots,j_L}= \tr(V_{j_1}^{i_1}\cdots V_{j_L}^{i_L})$ gives the same transformation as conjugating by $U_\textrm{chain}$. In the discussion below, we show that, after blocking sites, if we conjugate our integrable spin chain by \emph{any} MPU, then given the $\mathcal{R}$-matrix for the original integrable model, we can find another $\mathcal{R}$-matrix, depending on the MPU, that satisfies the Yang-Baxter equation. The new $\mathcal{R}$-matrix contains the original $\mathcal{R}$-matrix acting on a subspace of an enlarged virtual Hilbert space---see Figure \ref{fig:R'matrix}. On the other part of the enlarged space, there is a projection $\mathbb{I}\otimes \lvert r\rangle\langle l\rvert\otimes \mathbb{I}$ constructed from the eigenvectors of the MPU transfer matrix described in Section \ref{sec:MPU}. This projection means that this new $\mathcal{R}$-matrix does not satisfy a technical condition called regularity, and is moreover not invertible. However, we will show below that one can still deduce certain features of the model using this modified $\mathcal{R}$-matrix\footnote{Note that we do not exclude the possibility that there is some other natural modification of the $\mathcal{R}$-matrix.}, including the conserved charges. Thus we see that the underlying integrability is modified in a relatively straightforward way, depending on the original $\mathcal{R}$-matrix and the MPU. 
The Hamiltonian itself is in fact more closely related to another object called the $\mathcal{L}$-operator, introduced below. We will see that the $\mathcal{L}$-operator is acted on by the operators $V$ that form the MPU, illustrated in Figure \ref{fig:monodromy}.

\subsubsection{Classification of integrable models.} 
 In order to explore these ideas in greater depth, let us review some further details about the algebraic Bethe Ansatz. As mentioned, along with the $\mathcal{R}$-matrix we have the $\mathcal{L}$-operator. The  $\mathcal{L}$-operator satisfies the Yang-Baxter algebra depicted in Figure \ref{fig:YangBaxter}(b). Given an $\mathcal{L}$-operator that satisfies this algebra, one can generate further $\mathcal{L}$-operators satisfying the algebra. In this way one generates the Hamiltonian and other conserved charges \cite{Murg}.

Now, the fundamental solution to the Yang-Baxter algebra is given by a `rotated' $\mathcal{R}$-matrix. In particular, if we identify the horizontal (virtual) Hilbert space, and the vertical (physical) Hilbert space, then identifying the upper left and lower right indices of the $\mathcal{R}$-matrix as physical indices (rotating the picture by $\pi/4$) gives an $\mathcal{L}$-operator that solves \ref{fig:YangBaxter}(b). This is, however, not a unique solution to the Yang-Baxter algebra. 

\begin{figure}\scalebox{0.8}{
\includegraphics{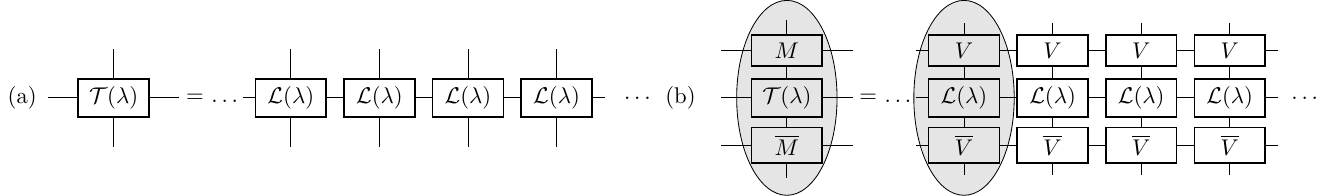}}
\caption{The monodromy matrix (a) before and (b) after conjugation by an MPU $M$ constructed from tensors $(V_j^i)^\alpha_{\beta}$. In (b) the shaded tensor on the left is $\mathcal{T}'(\lambda)$ while the shaded tensor on the right is $\mathcal{L}'(\lambda)$.}
\label{fig:monodromy}
\end{figure}

In general, the $\mathcal{L}$-operator is a tensor acting on a single site, with two physical indices and two uncontracted virtual indices. From $\mathcal{L}$-operators, $\mathcal{L}(\lambda)$, one can form the monodromy matrix\footnote{Typically one includes on-site rapidities $\xi_j$, i.e. $\mathcal{T}(\lambda;\xi_1,\dots,\xi_L)^{\alpha}_{\beta} = \mathcal{L}_1(\lambda-\xi_1)^{\alpha}_{\alpha_2}\mathcal{L}_2(\lambda-\xi_2)^{\alpha_2}_{\alpha_3}\cdots \mathcal{L}_L(\lambda-\xi_L)^{\alpha_{L}}_{\beta}$. These are helpful in certain intermediate calculations, then one can restore translation invariance by putting $\xi_j = \xi_0$ at the end \cite{Franchini}. For the purposes of our discussion we suppress these variables. In any case, our intention is to use the transformation to carry over known results rather than to rederive them from scratch.} \begin{align}\mathcal{T}(\lambda)^{\alpha}_{\beta} =\sum_{\alpha_2,\dots,\alpha_L} \mathcal{L}_1(\lambda)^{\alpha}_{\alpha_2}\mathcal{L}_2(\lambda)^{\alpha_2}_{\alpha_3}\cdots \mathcal{L}_L(\lambda)^{\alpha_{L}}_{\beta}.\end{align} This is an MPO on $L$ sites---depicted in Figure \ref{fig:monodromy}(a)---and is a tensor formed by matrix multiplication of the virtual indices of the $\mathcal{L}$ operators. 
$\mathcal{T}(\lambda)$ acts as a matrix on physical indices and  has two uncontracted virtual indices: by taking the trace we reach the transfer matrix $T(\lambda)$, where  $T(\lambda)$ has physical indices only. Transfer matrices at different values of the spectral parameter commute, $[T(\lambda),T(\mu)]=0$, and this means that the transfer matrix is the generating function of the conserved charges. In particular the canonical Hamiltonian of the integrable model is the logarithmic derivative of the transfer matrix at $\lambda =0$ \cite{Franchini}. 

Now let us address the question of classifying integrable models. One approach is to classify solutions to the Yang-Baxter equation, or $\mathcal{R}$-matrices, of some type \cite{Kulish1982}. For example the case corresponding to nearest-neighbour Hamiltonians was studied recently in \cite{deLeeuw2019,deLeeuw2020} for on-site Hilbert space $\mathbb{C}^2$ and $\mathbb{C}^4$ respectively (although it remains to apply Bethe Ansatz techniques to find the spectrum and eigenstates of some of these examples). By the discussion above, this would classify fundamental integrable models. In those works there is discussion of the fact that there are whole classes of equivalent spin chains that can be found by applying local basis transformations to the $\mathcal{R}$-matrix---this is equivalent to applying single-site basis transformations to the fundamental spin chain Hamiltonian. In \cite{Izergin1984} the following analogy with representation theory is drawn: fixing a particular $\mathcal{R}$-matrix is similar to fixing a particular group, while enumerating all monodromy matrices for a given $\mathcal{R}$-matrix is analogous to constructing representations of the group. Representations are reducible in general, so one can try to find the analogue of irreducible representations---these are the $\mathcal{L}$-operators studied in \cite{Izergin1984}. 

\begin{figure}[t]
\scalebox{0.75}{
\includegraphics{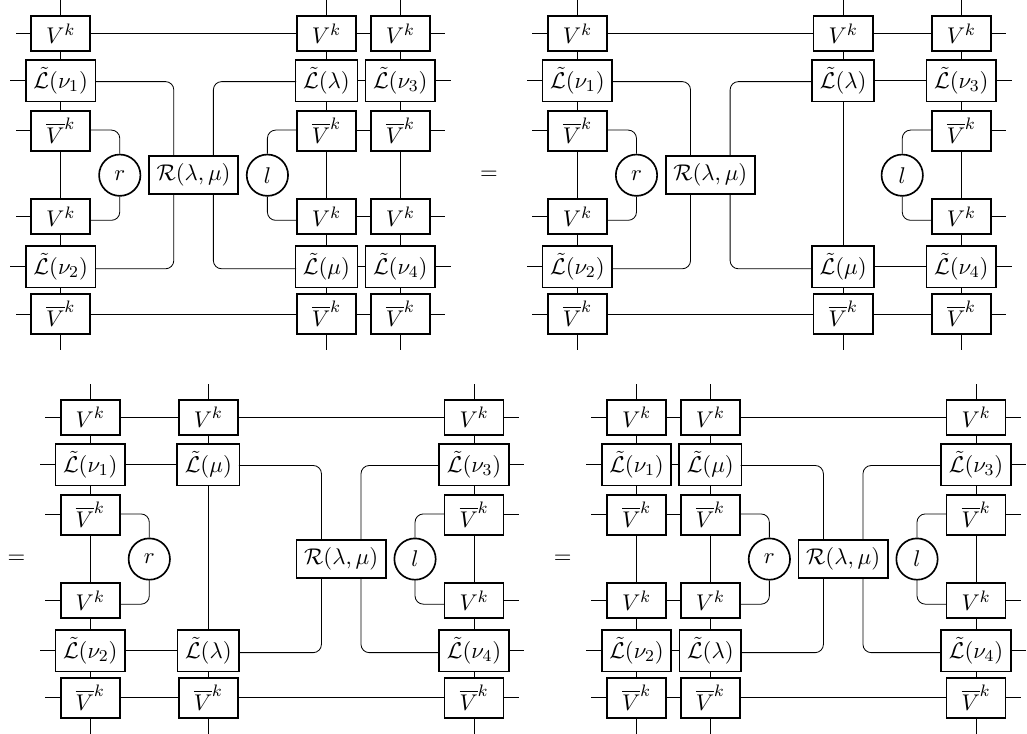}}
 \caption{A variant of the Yang-Baxter algebra for the $\mathcal{R}'$ matrix. Physical sites are blocked ($k$ times) so that the MPU satisfies the fixed-point equations. We can interpret the left-hand-side equalling the right-hand-side as the usual Yang-Baxter algebra holding whenever we contract virtual indices with additional (sufficiently blocked) MPU tensors such that we can use the pulling through condition. Indeed, the first equality is an application of the pulling through condition for the MPU, the second equality follows from the \emph{usual} Yang-Baxter algebra for the original $\mathcal{R}$-matrix, and then the third equality uses the pulling through condition again to give the analogue of the Yang-Baxter algebra.} 
 \label{fig:Fig8}
\end{figure} 

\begin{figure}[t]
\scalebox{0.75}{
\includegraphics{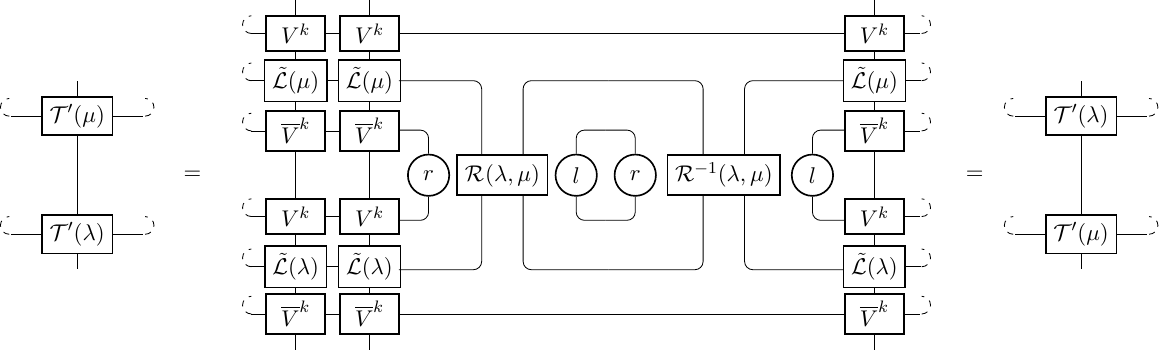}}
 \caption{A graphical proof that the transfer matrices commute for different values of the spectral parameter. The dashed lines indicate tracing over the virtual indices. In writing the first equality, we assume that the $\mathcal{R}$-matrix is invertible, and then use the fixed-point MPU equations to insert the projector $ \lvert r\rangle\langle l\rvert$. The modified Yang-Baxter algebra and the fixed-point MPU equation give the second equality.} 
 \label{fig:Fig9}
\end{figure}

In Figures \ref{fig:R'matrix} and \ref{fig:monodromy}(b)  we demonstrate the action of conjugation by an MPU on the various operators. Firstly, in our conjugated model we can define a new $\mathcal{L}$-operator, $\mathcal{L'}$, that includes the $V$ and $\overline{V}$ tensors that constitute the MPU (see the shaded tensor in Figure \ref{fig:monodromy}(b)) . The dimension of the physical indices of $\mathcal{L'}$ (corresponding to the dimension of the site of the spin chain) is unchanged, while the bond dimension of $\mathcal{L'}$ is the bond dimension of $\mathcal{L}$ multiplied by $\chi^2$, where $\chi$ is the bond dimension of the tensor $V$. We also define a blocked operator $\mathcal{\tilde{L}'}$, where we block $k$ physical sites so that the MPU reaches its fixed point. In Figure \ref{fig:Fig8}, we show that $\mathcal{\tilde{L}'}$ satisfies a modified Yang-Baxter algebra, with a modified $\mathcal{R}$-matrix: $\mathcal{R'}$; this is itself defined in Figure \ref{fig:R'matrix}. In Figure \ref{fig:Fig9} we show that despite this modification of the algebra, we can still use it to show the commutativity of the corresponding transfer matrices.

To summarise, we have shown that by applying MPU transformations to integrable models, we find a modified form of the underlying integrability. We have derived an underlying $\mathcal{R}$-matrix, where the dependence on the spectral parameter is through the $\mathcal{R}$-matrix of the original model only. This $\mathcal{R}$-matrix of the transformed model is not regular, and can be written as a tensor product of the original (assumed regular) $\mathcal{R}$-matrix and a trivial (but not regular) solution of the Yang-Baxter equation that is itself derived from the MPU. Moreover, we find a transformed $\mathcal{L}$-operator that satisfies a modified Yang-Baxter algebra. These transformed models fall outside the usual equivalence under basis transformations to the indices of the  $\mathcal{R}$-matrix; indeed this must be the case since the Hamiltonian is no longer nearest neighbour in general. We hence conclude that the essential features of the underlying integrability are unchanged, but our transformed models are not equivalent to the original model in the usual classification scheme. It would be most interesting to take this analysis further, for example by showing the existence of an $\mathcal{R}$-matrix and $\mathcal{L}$-operator that give  the transformed Hamilton and satisfy the unmodified Yang-Baxter algebra. One interesting point is whether such an  $\mathcal{L}$-operator would have a blocked physical index, and, if so, how this blocking relates to the range of the transformed Hamiltonian. 
 
\subsubsection{Relation to translation invariant Clifford transformations:}
We argued above that the Clifford transformations $U_\textrm{chain}$ in Table \ref{table:transformationstable} be written in an MPU form. This means that we can directly use the above analysis to see that the integrable spin chains in Section \ref{sec:summary} correspond to Bethe Ansatz solvable models. These models have an $\mathcal{R}$-matrix that is simply related to the XXZ  $\mathcal{R}$-matrix, and the $\mathcal{L}$-operator is not fundamental, but is related to the fundamental  $\mathcal{L}$-operator of the XXZ chain by conjugating with the appropriate MPU tensor. These operators do not satisfy the usual Yang-Baxter algebra, but a variant of this algebra that nevertheless allows us to prove commutativity of transfer matrices in the usual way.

The locality-non-preserving transformations do not fit directly into this analysis, as we do not have an MPU form. However, we can consider the system on open boundaries. Then the formalism of the algebraic Bethe Ansatz is more involved, but can still be described in terms of tensor networks \cite{Murg}. Moreover, the locality-non-preserving transformations that we consider can be written as unitary MPOs. It would be interesting to generalise the above analysis to the open boundary condition case to better understand the locality-non-preserving transformations discussed in Section \ref{sec:nonlocal}. We also note that the $\mathcal{L}$-operator and (implicitly) the $\mathcal{R}$-matrix for the bond-site transformed XYZ model (see also the model \eqref{eq:H''XXZ}) are given in \cite{Gombor21}. It would be interesting to connect their results with the methods of this paper.

\subsection{Free-fermion chains:}\label{sec:free} We now turn to free models. In this section we introduce a standard spin chain that can be analysed by transforming to a free-fermion model. Then we point out a construction, using Clifford group transformations, of further spin chain Hamiltonians that can be solved by free fermions. 

Now, focusing on translation invariant spin-$\sfrac{1}{2}$ chains, if we can find a transformation into fermionic modes such that the resulting Hamiltonian is quadratic, then this is a free-fermion chain\footnote{There are examples of chains that are free-fermion but cannot be solved by a Jordan-Wigner transformation---for discussion see \cite{Fendley2019,Elman21}.} \cite{Lieb1961}.  An example is the Ising model \begin{align}H_{\textrm{Ising}}=\frac{1}{2}\sum_{n \in \mathrm{sites}}\left(h Z_n- X_nX_{n+1}  \right),\end{align} and the Jordan-Wigner transformation takes us to a quadratic fermion model. In particular, define 
\begin{align}
\gamma_n = \prod_{m=1}^{n-1} Z_m X_n \qquad
\tilde \gamma_n = \prod_{m=1}^{n-1} Z_m Y_n, \label{eq:JordanWigner}
\end{align}
where $c_n = (\gamma_n+\rmi \tilde\gamma_n)/2$ is a canonical spinless fermion at site $n$, then $H_{\textrm{Ising}}= \frac{\rmi}{2}\sum_n\left( \tilde\gamma_n\gamma_{n+1} + h \tilde\gamma_n\gamma_{n}\right)$. This is a nearest neighbour model. An interesting class of longer-range spin chains are the generalised cluster models \cite{Suzuki71,Keating2004,Verresen17}:
\begin{align}\label{eq:H_free}
H_{\textrm{free}} = \frac{1}{2} \sum_{n\in\mathrm{sites}}\Bigg( t_0 Z_n-& t_1X_nX_{n+1} - t_2X_nZ_{n+1}X_{n+2} - t_3 X_nZ_{n+1}Z_{n+2}X_{n+3} - \dots \nonumber\\
&-t_{-1}Y_nY_{n+1}- t_{-2}Y_nZ_{n+1}Y_{n+2} - t_{-3} Y_nZ_{n+1}Z_{n+2}Y_{n+3}-\dots\Bigg),\end{align} 
where we can extend to longer range terms by including longer strings of $Z$ operators. Using the same Jordan-Wigner transformation \eqref{eq:JordanWigner}, this can be mapped into a quadratic fermion Hamiltonian with terms coupling sites beyond nearest neighbour: $H_{\textrm{fermion}}=\sfrac{\rmi}{2} \sum_{n,\alpha} t_\alpha \tilde\gamma_n \gamma_{n+\alpha}$. This can be solved similarly to the Ising model, giving the spectrum and eigenstates. Moreover, one can derive many results for the phase diagram, correlations and topological edge modes \cite{Barouch,deGottardi13,Verresen2018,Jones2019}. Note that these models have fermionic couplings only of the form $\rmi\tilde\gamma_n \gamma_m$, more generally one could also include couplings  $\rmi\tilde\gamma_n \tilde\gamma_m$. 

 Now, one can apply Clifford group transformations $U_\textrm{chain}$ to these spin chains and the transformed model will remain solvable by transforming to free fermions, and moreover we retain all of the mentioned results (appropriately transformed). The transformation $\tilde U_4$ of Table \ref{table:transformationstable}, for example, leads to the following transformations of terms from \eqref{eq:H_free} up to four-sites (one and two-site terms appear already in Section \ref{sec:summary}):\begin{align}
 X_nZ_{n+1}Z_{n+2}X_{n+3} &\rightarrow Z_{n-2} Y_{n-1} X_{n} Z_{n+1} Z_{n+2} X_{n+3} Y_{n+4} Z_{n+5},\nonumber\\
 X_nZ_{n+1}X_{n+2} &\rightarrow Z_{n-2} Y_{n-1} Y_{n} X_{n+1} Y_{n+2} Y_{n+3} Z_{n+4},\nonumber\\
 Y_nZ_{n+1}Y_{n+2}&\rightarrow -Z_{n-2} X_{n-1} Z_{n} X_{n+1} Z_{n+2} X_{n+3} Z_{n+4},\nonumber\\
 Y_nZ_{n+1}Z_{n+2}Y_{n+3} &\rightarrow Z_{n-2}X_{n-1}X_{n+4}Z_{n+5}.
 \end{align} 
One way to approach these free-fermion models is to transform the definitions of the Majorana fermions \eqref{eq:JordanWigner}:
\begin{align}
\gamma_n = \prod_{m=1}^{n-1} (UZ_mU^\dagger) (UX_nU^\dagger)\nonumber\\
\tilde \gamma_n = \prod_{m=1}^{n-1} (UZ_mU^\dagger) (UY_nU^\dagger),
\end{align}
and then inserting these definitions into the Hamiltonian $H_{\textrm{fermion}}$. A related approach is explored in \cite{Minami17}---there the starting point is any two Pauli operators that obey the commutation algebra of the Ising model terms $Z_n$ and $X_nX_{n+1}$ and then building a Jordan-Wigner transformation from them. By applying $U_\textrm{chain}$ to the operators $Z_n$ and $X_nX_{n+1}$ we preserve this algebra, so our results gives a different perspective on that work. We also mention that References \cite{Chapman2020,Ogura2020,Elman21} discuss the question of whether a particular spin chain Hamiltonian can be transformed into a free-fermion form. In these papers, the commutation relations between the terms in the Hamiltonian defines a graph and then graph-theoretic conditions for the existence of a free-fermion solution are given (see also earlier work on bond algebras \cite{Nussinov09}). Note that this graph depends on the basis of Pauli operators with which we expand the Hamiltonian. As pointed out in \cite{Chapman2020}, this graph is invariant under Clifford transformations, and so given a Hamiltonian of the form \eqref{eq:generalhamiltonian} the transformations discussed in this paper map between families of spin chains with the same underlying graph.

\section{Discussion}
In this paper, we have explained how to generate new families of integrable Hamiltonians using translation invariant Clifford transformations. As an example, we classified certain families of transformed XXZ models. These families were derived from a classification of translation invariant Clifford transformations that are built from $U\in\mathcal{C}_2$, as well as a classification of translation invariant Clifford transformations that take single-site Pauli operators to local operators on at most five sites. Finally, we put these results in context by discussing how they relate to usual classifications of integrable Hamiltonians. 

This is, however, far from a complete picture. We do not know how to determine whether or not two given Hamiltonians are related by a Clifford transformation. Understanding how to do this is an avenue of future research. As mentioned, there do exist interesting (conjectural) tests for integrability of spin chains \cite{Grabowski,Gombor21}. These conjectures are based on the existence of a conserved charge. The models we have considered that are related to the XXZ chain will naturally have a conserved charge corresponding to a certain operator in the XXZ chain. However, the second conjecture of \cite{Grabowski} relies on deriving this charge from a `boost operator', and since the models in Section \ref{sec:summary} are not nearest neighbour, the transformed boost operator will not be of the standard form given there. A generalisation to the case of medium-range models was given recently in \cite{Gombor21} and it would be interesting to understand whether our models satisfy the conjectures given in that paper. Furthermore, there are unresolved questions regarding the underlying integrable structures for the transformed models considered above.
Finally, as discussed above, the recent papers \cite{Chapman2020,Ogura2020,Elman21} analyse the question of when a spin chain can be solved by transforming to a free-fermion model.  It would certainly be of interest to find related results for integrable models beyond free fermions, such as those we have considered in this paper.

In terms of applications, we have already pointed out the appearance of Clifford transformations in the literature on SPT phases, as well as given a discussion of $H_0$ as a transition between SPT phases. As another example, one can take the critical Ising model 
\begin{align}
H_{\textrm{Ising}} = -\sum_{n \in \textrm{sites}} \Big (Z_nZ_{n+1} +X_n \Big) 
\end{align}
and using the transformation \eqref{eq:clustermapping} map it to 
\begin{align}
H_{\textrm{Cluster-Ising}} = -\sum_{n \in \textrm{sites}} \Big (Z_nZ_{n+1} +Z_{n-1}X_nZ_{n+1} \Big).
\end{align}
One notable feature of this model is that its low energy physics is described by a symmetry-enriched Ising CFT \cite{Verresen21}. The symmetry in this case is $\mathbb{Z}_2\times\mathbb{Z}_2^T$, generated by the spin-flip $\prod_n X_n$ and an antiunitary symmetry $T$ that acts as complex-conjugation (in particular, $X_n$ and $Z_n$ are real, while $Y_n$ is imaginary). In general, Clifford transformations will not preserve the symmetry of the Hamiltonian. A key point is that for a Clifford transformation on the whole chain that takes us from one symmetric Hamiltonian to another, the basic Clifford transformation that we use to construct it need not be symmetric. This property allows us to map between distinct symmetric phases as in the example above. Our methods give a general family of transformations that we could use to explore mapping standard integrable models to symmetry-enriched examples in a similar way. Conversely, given a Hamiltonian and a certain choice of symmetry group, one could classify the Clifford transformations considered above according to how they affect these symmetry properties. In Section \ref{sec:folded} we made connections to a recent work \cite{Tantivasadakarn2021} on SPT entanglers and pivot Hamiltonians. It would be interesting to understand more deeply these connections. For example, understanding the integrability properties and symmetry properties of pivoted integrable models would be most interesting. As mentioned, this could give new perspectives on the folded XXZ model. 

In this work we focus on integrable spin chains, including spin chains that can be solved by mapping into free-fermion chains. One may be directly interested in fermionic integrable models, for example, the Hubbard chain \cite{Lieb68,Korepin1997}. In that case, it would be natural to consider generalising our discussion above to include fermionic transformations. This would include Majorana translations $\gamma_n \rightarrow\gamma_{n+1}$ that do not have an analogue in the spin chain \cite{Fidkowski19}.  Moreover, in the fermionic case the equivalence between MPUs and QCAs that we discuss above no longer holds \cite{Piroli21}.

Finally, while we have focused our discussion on integrable models, the Clifford transformations $U_\textrm{chain}$ could of course be applied to any Hamiltonian. A motivation for restricting to integrable models was that we could use the many results about integrable models to analyse the resulting transformed model. For general models, it may still be of interest to understand which families of models are equivalent under the transformations that we have analysed here.

\section*{Acknowledgements}
We thank R. Verresen for a number of stimulating conversations as well as for insightful comments on the manuscript.
We are also grateful to P. Fendley, S. Piddock, L. Piroli, B. Pozsgay, D. Shepherd and R. Thorngren for helpful discussions and correspondence. We thank Y. Miao for valuable discussions and for identifying an issue with Section \ref{sec:Bethe} of the first version of this manuscript.
N. L. gratefully acknowledges support from the UK
Engineering and Physical Sciences Research Council
through Grants No. EP/R043957/1, No. EP/S005021/1,
and No. EP/T001062/1.

\bibliography{arxiv.bbl}{}

\appendix
\section{\texorpdfstring{$H_0$}{H0} is a transition between SPT phases}\label{app:H0}
In this appendix we will discuss nearby phases of the integrable model given in \eqref{eq:H_0}, which for convenience we repeat here, allowing for arbitrary (real) coupling constants:
\begin{align}
&H_0 = -\sum_{n \in \textrm{sites}} \Big(J_1 Z_{n-3} Y_{n-2}X_{n-1}Y_{n}Y_{n+1}X_{n+2}Y_{n+3}Z_{n+4}\nonumber\\&\qquad\qquad+J_2 Z_{n-3} Y_{n-2}Y_{n-1}Y_{n+2}Y_{n+3}Z_{n+4}+  J_3 Z_{n-1}Y_{n}Y_{n+1} Z_{n+2}\Big).
\end{align}
Recall also that $H_0=U_0H_{\mathrm{XYZ}}U_0^\dagger$ where $U_0=\tilde{U}_4\tilde{U}_2$ are defined in Table \ref{table:transformationstable} and $H_{\mathrm{XYZ}}$ is given by
\begin{align}
H_\mathrm{XYZ} = -\sum_{n \in \textrm{sites}} \Big(J_1 Y_{n}Y_{n+1} + J_2 X_nX_{n+1} + J_3 Z_nZ_{n+1}\Big) \label{eq:HXYZ}.
\end{align}
For $J_1=J_2=1$, $J_3=\Delta$ this is the XXZ model. 

Let us first discuss $H_\mathrm{XYZ}$, setting $J_3=0$ for convenience: then we have the XY model \cite{Barouch}. This model has a $\mathbb{Z}_2\times\mathbb{Z}_2$ symmetry generated by $P_X=\prod_n X_n$ and  $P_Y=\prod_n Y_n$ (we also have that $P_Z=\prod_n Z_n=P_XP_Y$ is a symmetry; where we assume the chain has length a multiple of four for convenience); as well as an anti-unitary $\mathbb{Z}_2^T$ time-reversal symmetry, $T$, that corresponds to complex conjugation.
Now, for $J_1>J_2$, the ground state spontaneously breaks $P_X$ and $T$ with order parameter $\langle Y_n \rangle$. Similarly for $J_2>J_1$, the ground state spontaneously breaks $P_Y$ with order parameter $\langle X_n \rangle$. More generally, for $\lvert J_3\rvert<J_2<J_1$ the nearby phases are the same \cite{Baxter72}. Hence, the XXZ model is a transition between an $X$-ordered ferromagnetic phase and $Y$-ordered ferromagnetic phase. 

We can make the same analysis for $H_0$. This model has the same symmetries as $H_\mathrm{XYZ}$. Firstly, for $J_1>J_2>\lvert J_3\rvert $, by applying the unitary $U_0$, we have symmetry-breaking with order parameter $\langle Z_{n-3}X_{n-2}Y_n X_{n+2}Z_{n+3} \rangle$. Hence, the ground state spontaneously breaks $P_X$ and $T$, but preserves $P_Y$ and $P_Z T$. In the fixed-point limit $J_1\rightarrow \infty$, we have the following:
\begin{align}
&\lvert \langle Z_{n-3}X_{n-2}Y_n X_{n+2}Z_{n+3} \rangle\rvert = 1 \Rightarrow \lvert \langle Z_1Y_2Y_3X_4 \left(\prod_{n=5}^M Y_n\right) X_{M+1}Y_{M+2}Y_{M+3}Z_{M+4} \rangle\rvert =1.
\end{align}
Hence, this correlator is a string-order parameter \cite{Pollmann12} for the symmetry string $P_Y$, and the end-point operator ($X_{M+1}Y_{M+2}Y_{M+3}Z_{M+4}$) has a non-trivial charge under the unbroken symmetry $P_ZT$. This means that it is a non-trivial SPT phase, and this is a stable property for all  $J_1>J_2>\lvert J_3\rvert $.

Secondly, for $J_2>J_1>\lvert J_3\rvert$, we have symmetry-breaking with order parameter \begin{align} 
\langle Z_{n-3}X_{n-2}Z_{n-1}Z_nZ_{n+1}X_{n+2}Z_{n +3} \rangle.\end{align} The ground state spontaneously breaks $P_X$ and $P_Y$, but preserves $P_Z$ and $T$. Again considering the fixed-point limit we have that\begin{align}
\lvert\langle Z_{n-3}X_{n-2}Z_{n-1}Z_nZ_{n+1}X_{n+2}Z_{n+3} \rangle \rvert&=1 \Rightarrow\nonumber\\
\lvert \langle Z_1Y_2X_3Y_4X_5~&\left(\prod_{n=6}^M Z_n\right)~ X_{M+1}Y_{M+2}X_{M+3}Y_{M+4}Z_{M+5} \rangle\rvert=1; \end{align}
giving a string order parameter for the phase. The end-point operator ($X_{M+1}Y_{M+2}X_{M+3}Y_{M+4}Z_{M+5}$) has trivial charge under the unbroken symmetry $P_Z T$. Hence, we see that $H_0$ ($J_1=J_2>\lvert J_3\rvert$) is at a transition between two different SPT phases.

Note that for the XXZ chain, the analogous string orders are simply $P_X$ and $P_Y$, which have trivial end-points and hence no non-trivial charges appear.

\section{Locality-non-preserving transformations}\label{app:nonlocal}
Here we prove the claims about the different locality-non-preserving transformations arising from two-site Clifford transformations, given in Section \ref{sec:twositeClifford}. Recall that we use the following notation for images of $X_n$ and $Z_n$ when conjugating by $U_{n,n+1}$:
\begin{align}
X_n &\rightarrow s_X  L_{n\phantom{+1}}^{(X)} R_{n+1}^{(X)},\quad Z_n \rightarrow s_Z L_{n\phantom{+1}}^{(Z)} R_{n+1}^{(Z)} \qquad \Big(\Rightarrow\quad Y_n \rightarrow s_Xs_Z \rmi L_{n\phantom{+1}}^{(X)}L_{n\phantom{+1}}^{(Z)}  R_{n+1}^{(X)}R_{n+1}^{(Z)}\Big),
\end{align}
where $s_X,s_Z\in\{\pm1\}$.
Above, we considered the case $R^{(X)} = R^{(Z)} = R$ which gave rise to local transformations for $R=Y$. If instead $R=X$ or $R=Z$ then we reach a transformation of the type (NL1) given in \eqref{eq:nonlocaltransform1}. More precisely, suppose $R=X$, then we must have (up to signs):
 \begin{align}X_n&\rightarrow   S_n X_{n+1} \qquad \tilde{P}_{n+1} \rightarrow P_n  Y_{n+1} \nonumber \\
Z_n &\rightarrow  Q_n X_{n+1} \qquad \tilde{Q}_{n+1} \rightarrow  P_n Z_{n+1}, \label{eq:nl1}\end{align}
where $P$ and $Q$ are anti-commuting single-site Pauli operators and $S:=\rmi P Q$. $\tilde P$ and $\tilde Q$ are also anti-commuting single-site Pauli operators, in applying $U_\textrm{chain}$ we can fix them to be $\tilde P=Y$ and $\tilde Q=Z$ by an initial choice of on-site basis. The string comes from repeated transformations of $X$. We see that under conjugation by $U_\textrm{chain}$ we have \eqref{eq:nonlocaltransform1} with boundary term $T=X$. Now, suppose that in \eqref{eq:nl1} we take $X_n \rightarrow -S_n X_{n-1}$. Then we will have an oscillatory sign, depending on site-index: $(-1)^n$. Other signs in \eqref{eq:nl1} will change overall signs, but not give this oscillatory behaviour, which comes from repeated transformations of $X$. Note that if we take $R=Z$, we would have $T=Z$; while, for completeness, the two-site Clifford:
\begin{align}Y_n&\rightarrow   S_n Y_{n+1} \qquad Y_{n+1} \rightarrow P_n  X_{n+1} \nonumber \\
Z_n &\rightarrow  Q_n Y_{n+1} \qquad Z_{n+1} \rightarrow  P_n Z_{n+1}, \end{align}
will give the same transformation \eqref{eq:nonlocaltransform1}, but with $T=Y$.

 We then need to analyse the cases where $R^{(X)} \neq R^{(Z)}$ and both are non-trivial, and so neither are equal to $R^{(Y)}=\rmi R^{(X)}R^{(Z)}$. Now, if $L^{(X)} =L^{(Z)} = \mathbb{I}$ then $U_\textrm{chain}$ is a swap transformation (up to an on-site change of basis) that is local in the bulk. Without loss of generality, we then have $L^{(X)} =L^{(Z)} \neq \mathbb{Id}$. Let us first consider $L^{(X)} =L^{(Z)}=R^{(Y)}$, this implies that the basic Clifford acts as:
 \begin{align}X_n&\rightarrow   R_{n\phantom{+1}}^{(Y)} R_{n+1}^{(X)} \qquad \tilde{P}_{n+1} \rightarrow   R_{n\phantom{+1}}^{(X)} R_{n+1}^{(Y)} \nonumber \\
Z_n &\rightarrow  R_{n\phantom{+1}}^{(Y)} R_{n+1}^{(Z)}\qquad \tilde Q_{n+1} \rightarrow  R_{n\phantom{+1}}^{(Z)} R_{n+1}^{(Y)}, \end{align}
where $\tilde P \neq \tilde Q \in \mathcal{P}\setminus \{\mathbb{I}\}$.
We then have three different cases that can be understood by\footnote{Other transformations fall into these three cases depending on whether the Pauli operator on site $n+1$ in the image is the same as the Pauli operator before the transformation. For example, $X_{n}\rightarrow Z_{n}Y_{n+1}$ and $Z_{n}\rightarrow Z_nX_{n+1}$ is of type (NL3) since the Pauli operator on site $n+1$ in each image is different to the initial Pauli operator on site $n$.  $X_{n}\rightarrow X_{n}Y_{n+1}$ and $Z_{n}\rightarrow X_nZ_{n+1}$ is of type (NL2) since exactly one Pauli operator ($Z$) has the same Pauli operator on site $n+1$ in its image.}:
\begin{align} 
\mathrm{(NL2)}\qquad &X_{n}\rightarrow s_XZ_{n}X_{n+1} \qquad   Z_{n}\rightarrow s_ZZ_nY_{n+1} \label{eq:nonlocaltransform2example2}\\
 \mathrm{(NL3)} \qquad&X_{n}\rightarrow s_XX_{n}Z_{n+1} 
 \qquad Z_{n}\rightarrow s_ZX_nY_{n+1} \label{eq:nonlocaltransform2example3}\\\
\mathrm{(NL4)} \qquad& X_n\rightarrow s_X Y_n X_{n+1} \qquad Z_n\rightarrow s_Z Y_nZ_{n+1}, \label{eq:nonlocaltransform2example4}
\end{align} 
the choice of signs $s_X, s_Z$ give different oscillatory terms (see Table \ref{table:oscillations}). The three different cases correspond to the remaining transformations given in Section \ref{sec:twositeClifford}. In particular, \eqref{eq:nonlocaltransform2example2} leads to \eqref{eq:nonlocaltransform2}, \eqref{eq:nonlocaltransform2example3} leads to \eqref{eq:nonlocaltransform3} and \eqref{eq:nonlocaltransform2example4} leads to \eqref{eq:nonlocaltransform4}. The oscillatory terms are determined by explicit calculation.

\begin{table*}[ht]
\begin{center}\begin{tabular}{|c|c|c|c|c|}\hline $(s_X,s_Z)$ & $(1,1)$ & $(1,-1)$ & $(-1,1)$ & $(-1,-1)$ \\ \hline Transformation \eqref{eq:nonlocaltransform2example2}& $\lfloor n/2\rfloor +b$ & $\lfloor n/2\rfloor +n+b$ & $b$ & $n +b$  \\\hline Transformation \eqref{eq:nonlocaltransform2example3}& $b $&$\lfloor n/3\rfloor+n +b$ & $\lfloor n/3\rfloor+\lfloor2n/3\rfloor+n +b $& $\lfloor2n/3\rfloor +b$ \\\hline Transformation \eqref{eq:nonlocaltransform2example4}& $b$ & $n+b$ & $n+b$ & $b$ 
\\\hline\end{tabular}
\vspace{0.15cm} 
\caption{Oscillatory terms that appear in non-local transformations depending on the signs $(s_X,s_Z)$. For the image of $X_n$ ($Z_n$), the oscillation will be $(-1)^{x(n)}$ ($(-1)^{z(n)}$) where $x(n)$ ($z(n)$) is the appropriate entry of the table. We have $b\in\{0,1\}$, where $b$ may be different for $x(n)$ and $z(n)$ and is fixed by the other signs in the basic Clifford transformation. }
\label{table:oscillations}
\end{center}
\end{table*}

Finally, we consider the other possibility which is, without loss of generality, that $L^{(X)} = L^{(Z)} = R^{(X)}$. Then:
\begin{align}X_n &\rightarrow   R_{n\phantom{+1}}^{(X)} R_{n+1}^{(X)} \qquad \tilde P_{n+1} \rightarrow   R_{n\phantom{+1}}^{(Y)} R_{n+1}^{(Y)} \nonumber \\
Z_n &\rightarrow  R_{n\phantom{+1}}^{(X)} R_{n+1}^{(Z)}\qquad \tilde Q_{n+1} \rightarrow  R_{n\phantom{+1}}^{(Z)} R_{n+1}^{(Y)}. \end{align}
This is similar to the previous case, the string depends on the choices of $ R^{(X)}$ and $R^{(Z)}$, and we are again led to the three cases (NL2)--(NL4). One can take, for example:
\begin{align} 
\mathrm{(NL2)}\qquad &X_{n}\rightarrow s_X X_{n}X_{n+1} \qquad   Z_{n}\rightarrow s_Z X_nY_{n+1} \\
 \mathrm{(NL3)} \qquad&X_{n}\rightarrow s_X Z_{n}Z_{n+1} 
 \qquad Z_{n}\rightarrow s_ZZ_nY_{n+1} \\\
\mathrm{(NL4)} \qquad& X_n\rightarrow s_X X_n X_{n+1} \qquad Z_n\rightarrow s_Z X_nZ_{n+1}.
\end{align} 
The oscillatory factors for these transformations are also given in Table \ref{table:oscillations}---the relevant entry depends on the corresponding transformation \eqref{eq:nonlocaltransform2example2}--\eqref{eq:nonlocaltransform2example4} according to the type of non-local string (NL2)--(NL4).

Note that this discussion shows that for all choices of basic Clifford transformation that is not one of the locality-preserving cases, we have a transformation of the form (NL1)--(NL4). We also have all transformations of this form (ignoring boundary terms) since we can apply an on-site basis change after the transformation---this is equivalent to taking a different basic Clifford transformation. 
\section{Constraints from commutation relations}\label{app:constraints}
In this appendix we prove certain results about translation invariant Clifford transformations. As explained above, the results for two- and four-site images follow directly from results in the literature on Clifford QCAs; we find it helpful to have a self-contained presentation here. In particular, we see explicitly the general rule that the images of $X_n$ and $Z_n$ must be symmetric under reflection about a site \cite{Schumacher2004,Schlingemann2008}.
\subsection{Two-site}\label{app:two-site}
Suppose that we have a translation invariant Clifford group transformation that maps $X_n$ to a two-site Pauli operator, without loss of generality say sites $n-1$ and $n$. Then we must have that $X_n \rightarrow \pm A_{n-1}A_n,$ and that $Z_n \rightarrow \pm   \prod_{ j} Q_j^{(j)}$ where this string is non-local. 

To show this, let $X_n \rightarrow \pm A_{n-1}B_n$. Then since $[X_n,X_{n+1}]=0$, we see that $[A_n,B_n] =0$. Hence, since we have a genuine two-site Pauli operator, we must have that $A_n = B_n$, i.e., $X_i \rightarrow \pm A_{n-1}A_n.$ Now, consider $Z_n \rightarrow \prod_{ j} Q_j^{(j)}$; this must anticommute with $A_{n-1}A_n$, hence exactly one of $Q^{(n-1)}$ and $Q^{(n)}$ must be a non-trivial Pauli, different from $A$. Let us say that $Q^{(n)}$ anticommutes with $A$. Now, \begin{align}[Z_{n-1},X_n] = 0 \Leftrightarrow [Q_{n-1}^{(n)}Q_{n}^{(n+1)}, A_{n-1}A_n]=0,\end{align} and so must have that $Q^{(n+1)}$ anticommutes with $A$. By repeating this reasoning we end up with a non-local string of non-trivial Pauli operators to the right. If we instead have that $Q^{(n-1)}$ anticommutes with $A$, we would get a non-local string to the left.
\subsection{Three-site}\label{app:three-site} 
Suppose that we have a translation invariant Clifford group transformation that takes
\begin{align}
&X_n \rightarrow A_{n-1}B_n C_{n+1}\nonumber\\
&Z_n \rightarrow A'_{m-1} B'_mC'_{m+1}. \label{eq:3site}
\end{align}
Since $[X_n, X_{n+2}] = 0$ we must have $[C,A]=0$. Hence either
$X_n \rightarrow A_{n-1}B_n A_{n+1}$ or $X_n \rightarrow B_n$. Similarly $[Z_n, Z_{n+2}] = 0$ means $Z_n \rightarrow A'_{m-1}B_m A'_{m+1}$ or $Z_n \rightarrow B'_m$. 
\begin{itemize}
\item If  $X_n \rightarrow B_n$ and $Z_n \rightarrow B'_m$ then clearly $m=n$ and $B\neq B'$.
\item If  $X_n \rightarrow A_{n-1}B_nA_{n+1}$ and $Z_n \rightarrow B'_m$, then $m=n$. (Since the image of $Z_n$ anticommutes with the image of $X_n$ we must have $n-1\leq m\leq n+1$. If, say, $m=n+1$ then since the image of $Z_n$ anticommutes with the image of $X_n$ it would also anticommute with the image of $X_{n+2}$, which is a contradiction.) Then we have $B' \neq B$, so that $X_n \rightarrow A_{n-1}B_nA_{n+1}$, $Z_n \rightarrow B'_n$  and $Y_n \rightarrow \pm A_{n-1}B''_nA_{n+1}$. Since $[Y_{n+1},X_n]=0$, we must have that $A\neq B$ and $A \neq B'$. 
\item If  $X_n \rightarrow A_{n-1}B_nA_{n+1}$ and $Z_n \rightarrow A'_{m-1}B'_m A'_{m+2}$, then again $m=n$. Any other option would mean that the image of $Z_n$ and the image of $Z_{n'}$ both anticommute with $X_n$, where $n' \in\{n\pm 2, n\pm 3\}$. Then $[Z_{n+2},X_n]=0$ implies that $A' =A$. This is now the same as the previous case with $Z$ and $Y$ swapped. 
\end{itemize}
Additional signs on the right-hand-side of \eqref{eq:3site} are analysed in the same way. Hence, we can conclude that the most general translation invariant Clifford group transformation that takes us to strings of length at most three is of the form:
\begin{align}
&P_n \rightarrow \pm S'_{n-1}P'_n S'_{n+1}\nonumber\\
&Q_n \rightarrow \pm S'_{n-1}Q'_nS'_{n+1}, \qquad   S'_n = \rmi  P'_n Q'_n.
\end{align}
\subsection{Four-site}\label{app:four-site}
Suppose that we have a translation invariant Clifford group transformation that takes
\begin{align}
    X_n \rightarrow A_{n-1}B_n C_{n+1}D_{n+2}.
\end{align}
Then we must have that
\begin{align}
&X_n \rightarrow A_{n-1}B_n B_{n+1}A_{n+2}.\nonumber\\
\end{align}
and the image of $Z_n$ is infinite.

To derive the possible images of $X_n$, note that  $[X_n, X_{n+3}] = 0$ means that $A=D$. Then $[X_n, X_{n+1}] = [X_n, X_{n+2}]= 0$ gives us that $B=C$. To then show that the image of $Z_n$ is infinite, there are two cases to consider, $[B,A]=0$ and $\{B,A\}=0$. Let us consider the first case, then either $B=A$ or $B=\Id$. If $X_n \rightarrow A_{n-1}A_{n+2}$ then we can proceed as in the two-site case. The image of $Z_n$ must contain a Pauli operator either on site $n-1$ or on site $n+2$ that anticommutes with $A$, then since the image of $Z_n$ commutes with $A_{m-1} A_{m+2}$ for all $m\neq n$, the image of $Z_n$ must be an infinite string to either the left or the right. If $X_n \rightarrow A_{n-1}A_n A_{n+1}A_{n+2}$, letting $Z_n = \dots P^{(n-1)}_{n-1}P^{(n)}_n P^{(n+1)}_{n+1}P^{(n+2)}_{n+2}\dots $ we have that either one or three of $\{P^{(n-1)}, \dots, P^{(n+2)}\}$ anticommute with $A$. In the case that only $P^{(n-1)}$ ($P^{(n+2)}$) anticommutes with $A$, then the image of $Z_n$ is infinite to the left (right), otherwise it is infinite in both directions. This follows easily from the commutativity with images of $X_m$ for $m\neq n$.

For the case $\{B,A\}=0$ we use a different approach. In that case, 
\begin{align}
X_n \rightarrow A_{n-1}B_n B_{n+1}A_{n+2}  .
\end{align}
Let us call our four-site Clifford transformation $U_4$; we know from Appendix \ref{app:three-site} that we can find a translation invariant Clifford group transformation $U_3$ such that $A_n \rightarrow A_n $ and $B_n \rightarrow A_{n-1}B_n A_{n+1}$. Now, conjugating by $U_3U_4$ will take $X_n \rightarrow B_{n}B_{n+1}$, and will take $Z_n \rightarrow U_3U_4Z_nU^\dagger_4U^\dagger_3$. Since $U_3$ maps single-site operators to at most three-site operators, if $U_4Z_nU^\dagger_4$ is a finite string then $U_3U_4$ contradicts the result of Appendix \ref{app:two-site}. Hence, the image of $Z_n$ must be an infinite string.
\subsection{Five-site}
Suppose that we have a translation invariant Clifford group transformation that takes
\begin{align}
&X_n \rightarrow A_{n-2}B_{n-1}C_n D_{n+1}E_{n+2}\nonumber\\
&Z_n \rightarrow A'_{m-2}B'_{m-1}C'_m D'_{m+1}E'_{m+2}. 
\end{align}
Let us first consider the case where none of these operators are the identity. Then $[X_n, X_{n+4}] = [Z_n, Z_{n+4}] =0$ gives $A=E$ and $A'=E'$. That is,
\begin{align}
&X_n \rightarrow A_{n-2}B_{n-1}C_n D_{n+1}A_{n+2}\nonumber\\
&Z_n \rightarrow A'_{m-2}B'_{m-1}C'_m D'_{m+1}A'_{m+2}. 
\end{align}
Now, $[X_n, X_{n+3}]=0$ tell us that either $B= D = A$ or that both $B$ and $D$ are distinct from $A$. In the second case, $[X_n, X_{n+2}]=0$ gives us that $B=D$. The same considerations apply to the image of $Z$. Since both images are symmetric about the middle site, if $m\neq n$ we have an inconsistency: if $m=n+k$ then the image of $Z_{n-2k}$ will also anticommute with the image of $X_n$. Hence we have that the image of $X_n$ is either $A_{n-2}B_{n-1}C_n B_{n+1}A_{n+2}$ or $A_{n-2}A_{n-1}C_n A_{n+1}A_{n+2}$, while the image of $Z_n$ is either $A'_{n-2}B'_{n-1}C'_n B'_{n+1}A'_{n+2}$ or $A'_{n-2}A'_{n-1}C'_n A'_{n+1}A'_{n+2}$.
In all cases, $[X_n,Z_{n+4}]=0$ implies $A'=A$. Considering also $[X_n,Z_{n+3}]=0$, $[X_n,Z_{n+2}]=0$ and $\{X_n,Z_n\}=0$ give us two options
\begin{itemize}
\item The first case is:
\begin{align}
&X_n \rightarrow A_{n-2}A_{n-1}C_n A_{n+1}A_{n+2}\nonumber\\
&Z_n \rightarrow A_{n-2}A_{n-1}C'_n A_{n+1}A_{n+2}                 
\end{align}
 where $ C_n\neq C'_n ~\textrm{and}~ A_n=\rmi C_nC_n'$. 
\item
The second case is \begin{align}
&X_n \rightarrow A_{n-2}B_{n-1}C_n B_{n+1}A_{n+2}\nonumber\\
&Z_n \rightarrow A_{n-2}B'_{n-1}C'_n B'_{n+1}A_{n+2}\label{eq:5site2} \end{align}
with $A, B, C, C' \in \mathcal{P}$ such that: $ B\neq A, B'\neq A, C\neq C'$. Now, if $B'=B$ then $C\neq A$ and $C'\neq A$, but also $C\neq B$ and $C'\neq B$ which is inconsistent. Hence $B'\neq B$, and either $C=A$ or $C'=A$. If $C=A$, then $C' = B'$, otherwise if $C'=A$ then $C=B$. \end{itemize}
An example of this case is:
 \begin{align}
&X_n \rightarrow  X_{n-2}Z_{n-1}Z_n Z_{n+1}X_{n+2}\nonumber\\
&Z_n \rightarrow X_{n-2}Y_{n-1}X_n Y_{n+1}X_{n+2}\nonumber\\
&Y_n \rightarrow  X_{n-1} Y_n X_{n-2} \label{eq:5siteexample1} . \end{align}

Now let us consider the five-site case where we allow some operators to be the identity. Due to Appendix \ref{app:two-site} and \ref{app:four-site} we can take it that $X_n$ maps to a five-site, three-site or one-site operator. First let
\begin{align}
&X_n \rightarrow A_{n-2}B_{n-1}C_n D_{n+1}A_{n+2}\nonumber
\end{align}
where $A\neq\Id$. We can immediately exclude cases where each of $B,C,D \in \{\mathbb{I},A\}$, since, similarly to Appendix \ref{app:four-site}, one could then argue that the image of $Z_n$ is infinite.
Since the images of different $X_n$ commute, one can show that there are the following possibilities that include at least one identity and one operator not equal to $A$:
 \begin{align}
&X_n \rightarrow A_{n-2}C_n A_{n+2} \qquad\qquad \quad C\neq A\label{eq:X5site}\\
&X_n \rightarrow A_{n-2}B_{n-1} B_{n+1}A_{n+2} \qquad B\neq A\label{eq:X5site2}.
\end{align}
Let $U_5$ be any transformation that acts as \eqref{eq:X5site2}, and using $U_3$ of Appendix \ref{app:four-site}, we have that $X_i \rightarrow B_{n-1} B_{n+1}$ under conjugation by $U_3 U_5$. Clearly any image of $Z_n$ that anticommutes with that image and commutes with the images of all other $X_m$ will be infinite, and so the image of $Z_n$ under $U_5$ must be infinite. We can hence ignore transformations of the form \eqref{eq:X5site2}. 

 For the other possibility, \eqref{eq:X5site}, we can find images of $Z_n$ over five sites. Firstly, the image of $Z_n$ must be either one-site, three-site or five-site according to Appendix \ref{app:two-site} and \ref{app:four-site}. If the image of $Z_n$ is one site, the only consistent possibility is $Z_n \rightarrow A_n$. If the image of $Z_n$ is three-site, then we must have $Z_n\rightarrow A'_{m-1}B'_mA'_{m+1}$. Then the form of \eqref{eq:X5site} means the only consistent choice is $A'=\Id$. Finally if the image of $Z_n$ is five site, we have $Z_n \rightarrow A'_{m-2}B'_{m-1}C'_m B'_{m+1}A'_{m+2}$. Since only one image of $Z_n$ can anticommute with \eqref{eq:X5site}, we must have $A'=A$, and that $B'\in \{A,\Id\}$. 
 Then, by symmetry, we must have $m=n$ with $\{C',C\} =0$ and $C' \neq A$. By considering $[X_{n \pm1},Z_n]=0$ we have that $B'=\Id$. Note that in this final case, $Y_n \rightarrow \pm A_n$, so all five-site possibilities involving identity reduce to the same case. 
 
For completeness we also consider the cases where the image of $X_n$ is a three-site, or a one-site operator. Excluding cases treated in earlier appendices, we may assume that the image of $Z_n$ is a five-site operator. We now show that all such cases reduce to previously considered five-site transformations, up to some relabelling on the left-hand-side. If $X_n \rightarrow P_n$, a one-site operator, then it is easy to see from the commutation relations that $Z_n \rightarrow A_{n-2} B_{n-1} C_n B_{n+1} A_{n+2}$, where $C \neq A$ and where $[B,A] =0$. These possibilities have already appeared above. If the image of $X_n$ is a three-site operator then we have either  $X_n \rightarrow A_{n-1}A_{n+1}$ or $X_n \rightarrow A_{n-1}B_nA_{n+1}$. The first option leads to an infinite image of $Z_n$. The second option must have $B\neq A$, otherwise again we have an infinite image of $Z_n$. In particular, we have that $A\neq B$, $A,B\neq \Id$ and $Z_n \rightarrow A_{n-2} B'_{n-1} C_n B'_{n+1} A_{n+2}$. We have already seen that $B'=\Id \Rightarrow B=\Id$. If $B'\neq \Id$, then in all consistent cases we reach an example of the form \eqref{eq:5site2}.

 In conclusion, we have proved that all locality-preserving translation-invariant Clifford transformations that takes single-site Pauli operators to at most five-site Pauli operators must be of the form (L1)--(L6) given in Sections \ref{sec:XXZfamilies} and \ref{sec:generalisations}.
 
\end{document}